\documentclass{desyproc}
\usepackage{graphicx}
\usepackage{psfrag}
\usepackage{epstopdf}
\usepackage{epsfig}
\usepackage {wrapfig} 
\usepackage[dvipsnames]{xcolor}
\usepackage{color}
\usepackage{subfigure}
\usepackage{bm}

\newcommand{\ba}{\begin{array}}
\newcommand{\ea}{\end{array}}
\newcommand{\vx}{{\bar x}}
\newcommand{\vp}{{\bar p}}

\newcommand{\wvp}{\bar P}

\newcommand{\ms}{\mathstrut}
\newcommand{\ds}{\displaystyle}

\newcommand{\be}{\begin{equation}}
\newcommand{\ee}{\end{equation}}
\newcommand{\bea}{\begin{eqnarray}}
\newcommand{\eea}{\end{eqnarray}}

\begin{document}
\title{Particle Production in Strong Time-dependent Fields}

\author{{\slshape D. B. Blaschke$^{1,2,3}$, S.~A.~Smolyansky$^4$, A.~Panferov$^4$, 
L.~Juchnowski$^1$}\\[1ex]
$^1$Institute of Theoretical Physics, University of Wroclaw, 50-204 Wroclaw, Poland\\
$^2$Bogoliubov Laboratory for Theoretical Physics, JINR,
RU - 141980 Dubna, Russia\\
$^3$National Research Nuclear University (MEPhI), RU - 115409 Moscow, Russia\\
$^4$Saratov State University, RU - 410026 Saratov, Russia }

\contribID{xy}

\confID{999}  
\desyproc{DESY-PROC-2099-01}
\acronym{VIP2010} 
\doi  

\maketitle

\begin{abstract}
In these lecture notes we give an introduction to the kinetic equation approach to pair production form the vacuum in strong, time-dependent external fields (dynamical Schwinger process).
We first give a derivation of the kinetic equation with the source term for the case of fermions starting from the Dirac equation and for bosons from the Klein-Gordon equation. 
In a second part we discuss the application of the approach to the situation of external field pulses as single-sheeted functions of time (like the Sauter-pulse) and as multi-sheeted functions approximating the 
situation in the focal point of counter-propagating laser beams. 
Special emphasis is on the discussion of the time evolution of the system that exhibits the characteristics of a field-induced phase transition for which we discuss the behaviour of the entropy and particle density of the system. 
We give an outlook to applications of the approach in describing particle production in strong fields formed in particle and nuclear collisions. 
\end{abstract}

\section{Introduction}

Vacuum $e^+e^-$ pair creation by a classical electric field is a longstanding prediction in QED 
\cite{Sauter:1931zz,Heisenberg:1935qt,Schwinger:1951nm}.
A complete theoretical description of the effect exists 
\cite{Grib:1994,Nikishov:1969tt,Narozhnyi:1970uv,Greiner:1985ce,Fradkin:1991zq}, but there is still
no experimental verification. 
The main obstacle is the high value of the critical electric field strength for pair creation;
viz., $E_c = m^2/e = 1.3 \times 10^{16}$ V/cm for the electron-positron case.
According to the so-called Schwinger formula \cite{Schwinger:1951nm}, the pair creation rate in a 
constant electric field,
\bea
\label{Schwinger}
S^{\rm cl} = \frac{e^2 \ E^2}{4\pi^3} \exp\left(-\frac{\pi m^2}{|e E|}\right)~,
\eea
is suppressed exponentially when $E \ll E_c$.
However, a very different situation occurs when the field
acts only in a finite time interval (dynamical Schwinger effect) 
\cite{Grib:1994,Schmidt:1998vi,Blaschke:2013ip,Popov:2002}.
In this case, the Schwinger formula, as well as its analog for a monochromatic field 
(Brezin - Itzykson formula \cite{Brezin:1970xf}), become inapplicable in the weak
field regime.

A few examples have been discussed of physical situations where the Schwinger effect could occur despite the high critical field strength; e.g., relativistic heavy ion collisions 
\cite{Casher:1978wy,Bialas:1984wv,Ryblewski:2013eja}
and focused laser pulses \cite{Richards:1959}.
Since the Schwinger effect is non-perturbative and it
requires an exact solution of the dynamical equations it
is customary to approximate the complicated structure of
a real laser field by a spatially uniform time-dependent
electric field. 
According to different estimates 
\cite{Popov:2002,Brezin:1970xf,Casher:1978wy,Richards:1959,Bunkin:1969if,Bulanov:2003aj,Bulanov:2004de}
the effect of vacuum pair creation is unlikely to be observable
with presently available laser parameters.
However, recent developments in laser technology, in
particular the invention of the chirped pulse amplification
method, have resulted in a huge increase in the light
intensity at the laser focal spot 
\cite{Mourou:1998,Mourou:2006zz}. 
On this basis the European Extreme Light Infrastructure (ELI) project
will be developed in order to provide radiation beams of
femto- to atto-second duration in the deeply relativistic
regime, exceeding intensities of $10^{25}$ W/cm$^2$
\cite{Dunne:2008kc,Heinzl:2008an}. 
On the other hand, construction of X-ray free electron lasers
XFEL \cite{Ringwald:2001ib} based on the SASE principle is underway
at DESY Hamburg. 
Thus an experimental verification of the Schwinger effect is coming within reach.

Under conditions of short duration pulses time-dependent effects become important. 
Therefore in our works 
\cite{Schmidt:1998vi,Blaschke:2005hs,Blaschke:2008wf,Blaschke:2011is,Blaschke:2013ip,Otto:2015gla}
we have developed a kinetic equation approach,
which allows us to consider the dynamics of the vacuum
pair creation process while accounting properly for
the initial conditions \cite{Schmidt:1998vi,Pervushin:2003uu}. 
Compared to alternative treatments, this approach is essentially nonperturbative
and contains new dynamical aspects, such as longitudinal
momentum dependence in the distribution functions and non-Markovian character of the time evolution
\cite{Schmidt:1998zh}. 
It also takes into account the effects of field switching and particle statistics 
\cite{Schmidt:1998zh,Bloch:1999eu,Prozorkevich:2004yp}. 
This approach has been applied
already to the periodical field case \cite{Roberts:2002py} with near-critical
values of the field strength and X-ray frequencies. In
particular, it was shown that there is an accumulation
effect when the intensity of the field is about half critical:
the average density of pairs grows steadily with increasing
number of field periods. 
The method \cite{Schmidt:1998vi} also found
application in describing the pre-equilibrium evolution of
a quark-gluon plasma produced in ultrarelativistic heavy
ion collisions at RHIC and LHC 
\cite{Schmidt:1998zh,Bloch:1999eu,Prozorkevich:2004yp}. 

A characteristic feature of the kinetic approach is the possibility to describe the evolution of the 
particle distribution functions during all stages of the external field evolution.
In the second main part of these lectures we will investigate this aspect more in detail and 
elucidate that the process of pair creation in a strong, time-dependent external field can be viewed as a
field induced phase transition (FIPT) \cite{Smolyansky:2016gmp,Panferov_2017_2}.
We introduce the concept of an order parameter and describe its evolution through three stages, 
the initial quasiparticle (QEPP) stage, the transient stage and the final residual (REPP) stage of the 
electron-positron plasma (EPP) created in the external field.
We suggest that these features are rather universal and appear qualitatively similar in physical systems 
of different nature.  
In concluding these lectures we discuss the question related with the observation that the spectra 
of hadrons in ultrarelativistic heavy-ion collisions appear thermal despite the fact that the production 
mechanism by the Schwinger process would entail non-thermal spectra and time scales between creation and freeze-out are too short for equilibration by multiple collisions. 

\section{Dynamics of pair creation }
\subsection{Creation of fermion pairs}
This section contains the derivation of a kinetic equation for the fermion-antifermion pair production. 
For the description of $e^+e^-$ production in an electric field we start from the QED Lagrangian
\bea\label{1}
{\cal L} = {\bar \psi} i\gamma^\mu(\partial_\mu+ieA_\mu)\psi - m{\bar
\psi}\psi-\frac{1}{4}F_{\mu\nu}F^{\mu\nu}\,,
\eea
where $F^{\mu\nu}$ is the field strength, the metric is taken as $g^{\mu\nu} = {\rm diag}(1,-1,-1,-1)$ and for the $\gamma$-- matrices we use the conventional definition.
The equation of motion for the case of $e^{+}e^{-}$ production in an external classical  electric field is given by the Dirac equation
\bea\label{2}
(i\gamma^\mu\partial_\mu-e\gamma^\mu A_\mu-m)\psi(x)=0\,.
\eea
where $A_\mu$ denote the vector potential in Hamiltonian gauge
\begin{equation}
A_\mu=(0,0,0,A(t))
\end{equation}
The resulting electric field 
\bea\label{20}
E(t)=-{\dot A}\ms(t)=-dA(t)/dt 
\eea
is homogeneous in space and time-dependent. There is no magnetic component because 
\begin{equation}
\label{Bzero}
B = \nabla \times A = 0
\end{equation}
The scenario with purely electrical field described by (\ref{20}) and (\ref{Bzero}) can be realized in laboratory by two counter-propagating laser beams forming a standing wave. 
It is worth noticing that particle production occurs only if one of two Lorentz invariants
\begin{equation}
\label{lorentzS}
\frac{1}{4}F_{\mu\nu}F^{\mu\nu} = \frac{1}{2}(E^2 + B^2),
\end{equation}
\begin{equation}
\frac{1}{4}F_{\mu\nu}\tilde{F} ^{\mu\nu} = E\dot B
\end{equation}
is non-zero. For the present calculations (\ref{lorentzS}) has a non-zero value.

Since the Schwinger mechanism is not perturbative  one needs to find an exact solution. 
The first step is the introduction of an ansatz for the spinor field
\bea
\label{10} 
\psi^{(\pm)}_{\vp r}(x)& =& 
\bigg[i\gamma^0\partial_0+\gamma^kp_k-e\gamma^3A(t)+m\bigg] \
\chi^{(\pm)}(\vp,t) \ R_r \ {\rm e}^{i\vp\bar x},
\eea
where $k=1,2,3$ and $(\pm)$ denotes eigenstates with  positiv/negative frequencies.
\be
\label{30}
R_1 =\left(\begin{array}{r}0\\1\\0\\-1\end{array}\right)\,\,,\hspace{4cm}
R_2 =\left(\begin{array}{r}1\\0\\-1\\0\end{array}\right)\,\,\, ,
\ee
are eigenvectors of the matrix  $\gamma^0\gamma^3$ so that $R^+_r R_s = 2\delta _{rs}$.
Substitution of (\ref{10}) to (\ref{2}) gives the parametric oscillator-type equation
\bea
\label{40} {\ddot
\chi}^{(\pm)}(\vp,t)&=&-\bigg(\omega^2(\vp,t)+ie{\dot
A}(t)\bigg) \ \chi^{(\pm)}(\vp,t)\,\,.
\eea
The quasi-particle energy  $\omega(\vp,t)$, the transverse energy $\varepsilon_\perp$ and the
longitudinal quasi-particle momentum are defined as 
\begin{eqnarray}
\omega(\vp, t) &=& \sqrt{\varepsilon_\perp^2 (\vp_\perp) + P^2(\vp_\parallel, t)} ,
\label{eq:energy}
\\
\varepsilon_\perp (\vp_\perp)&=& \sqrt{m^2 + \vp^2_\bot},
\label{eq:energy_perp}
\\
 P(\vp_\parallel, t) &=& \vp_\parallel -eA(t),
\label{eq:p-long}
\end{eqnarray}
The system of the spinor functions (\ref{10}) is complete and orthonormalized so the field operators $\psi(x)$, ${\bar \psi}(x)$ can be
decomposed in the spinor functions (\ref{10})  as follows:
\be\label{70}
\psi(x)=\sum\limits_{r,\vp}
\bigg[\psi^{(-)}_{\vp r}(x)\ b_{\vp r}(t_0) +
\psi^{(+)}_{\vp r}(x) \ d^+_{-\vp r}(t_0)
\bigg] \,\,,  
\ee
where $b_{\vp r}(t_0),b^+_{\vp r}(t_0)$, $d_{\vp r}(t_0),d^+_{\vp r}(t_0)$ 
are creation and annihilation operators of electrons and positrons. 
They act on the vacuum in-state $|0_{\rm in}\rangle$ at the initial time $t=t_0$, and obey the anti-commutation relations
\be
\label{71} 
\{b_{\vp r}(t_0),b^+_{\vp ' r'}(t_0)\}=\{d_{\vp r}(t_0),d^+_{\vp ' r'}(t_0)\}= \delta_{rr'} \ \delta_{\vp \vp '} \, . 
\ee
In general the evolution of a relativistic system affects the vacuum state and 
mixes states with negative $\psi^{(-)}_{\vp r}(x)$ and positive $\psi^{(+)}_{\vp r}(x)$ energies. 
As a result  in the Hamiltonian appear non-diagonal terms that are responsible for pair creation. 
For the Hamiltonian corresponding to the Dirac equation (\ref{2}) in a homogeneous electric field the diagonalization is achieved by a  time-dependent Bogoliubov  transformation
%
\bea
\label{90}
b_{\vp r}(t) &=&\alpha_\vp (t)\ b_{\vp r}(t_0)+ \beta_\vp (t)\ d^+_{-\vp r}(t_0)\ , \\ 
d_{\vp r}(t)&=&\alpha_{-\vp} (t) d_{\vp r}(t_0) - \beta_{-\vp}(t) b^+_{-\vp r}(t_0)
\label{91}
\eea
with the normalization condition
\be
\label{110}
|\alpha_\vp(t)|^2+|\beta_\vp(t)|^2=1\,\,.  
\ee
Thus the new  operators $b_{\vp r}(t)$ and $d_{\vp r}(t)$ describe quasiparticles
 at the time $t$ with the instantaneous vacuum $|0_t \rangle$. 
 Clearly, the operator system $b(t_0),b^+(t_0);d(t_0),d^+(t_0)$ is unitary
non-equivalent to the system $b(t), b^+(t); d(t), d^+(t)$.
The application of the Bogoliubov  transformation to
Eq.~(\ref{70}) gives the new representation of the field operators
\bea
\label{120} 
\psi(x)&=&\sum_{r,\vp}\bigg[\Psi^{(-)}_{\vp r}(x) \ b_{\vp r}(t) + \Psi^{(+)}_{\vp r}(x) \ d^+_{-\vp r}(t)\bigg]\,\,.  \eea 
The correspondence between the new $\Psi^{(\pm)}_{\vp r}(x)$ and the former $\psi^{(\pm)}_{\vp r}(x)$ spinor functions is defined by a canonical transformation 
 \bea
\label{130}
\psi^{(-)}_{\vp r}(x)&=&\alpha_\vp(t) \ \Psi^{(-)}_{\vp r}(x) - \beta^*_\vp(t) \ \Psi^{(+)}_{\vp r}(x)\,\,,\\ 
\psi^{(+)}_{\vp r}(x)&=&\alpha^*_\vp(t) \ \Psi^{(+)}_{\vp r} (x)+ \beta_\vp(t) \ \Psi^{(-)}_{\vp r}(x)\,\,.
\label{131}
\eea
Therefore it is justified to assume that the functions $\Psi^{(\pm)}_
{\vp r}$ have a spin structure  similar to that of $\psi^{(+)}_{\vp r}$ in Eq.~(\ref{10}),
\be
\label{150}
\Psi^{(\pm)}_{\vp r}(x)=  
\bigg[i\gamma^0\partial_0+\gamma^kp_k-e\gamma^3A(t)+m\bigg]\phi^{(\pm)}_{\vp}(x) \ R_r \
{\rm e}^{\pm i\Theta(t)} {\rm e}^{i\vp\vx},
\ee
where the dynamical phase is defined as
\be
\label{160}
\Theta(\vp,t) = \int^t_{t_0}dt'\omega(\vp,t')\,\,.
\ee
and $\phi^{(\pm)}_{\vp}$ are yet unknown functions. 
The substitution of Eq.~(\ref{150}) into Eqs. ~(\ref{130}) and (\ref{131}) leads to the relations
\bea
\label{170}
\chi ^{(-)}(\vp,t)&=&\alpha_\vp(t) \ \phi_\vp^{(-)}(t) \ {\rm
e}^{-i\Theta(\vp,t)}-\beta^*_\vp(t) \ \phi^{(+)}_\vp(t) \ {\rm
e}^{i\Theta(\vp,t)}\,\,,\\
\chi^{(+)}(\vp,t)&=&\alpha^*_\vp(t) \ \phi_\vp^{(+)}(t) \ {\rm
e}^{i\Theta(\vp,t)}+\beta_\vp(t) \ \phi^{(-)}_\vp(t) \
{\rm e}^{-i\Theta(\vp,t)}\,\,.
\label{171} 
\eea
On the other hand $\chi^{(\pm)}$ are solutions of Eq.~(\ref{40}). Their behavior is directly related to the asymptotics of the vector potential.
At $t_0=t\rightarrow -\infty$ we have $A(t_0) = 0$ so that

\be
\label{60}
\chi^{(\pm)}(\vp,t) 
\sim\exp{\big(\pm i\omega_0(\vp)\,t\big)}\,\,.
\ee

Now,  according to the Lagrange method, we can use Eqs.~(\ref{40}) and (\ref{60})  to 
introduce additional constraints on $\chi^{(\pm)}$


\bea
\label{190}
 {\dot \chi}^{(-)}(\vp,t)&=&-i\omega(\vp,t)\ \bigg[\alpha_\vp(t) \
\phi_\vp^{(-)}(t) \ {\rm e}^{-i \Theta(\vp,t)}+ \beta^*_\vp(t) \
\phi^{(+)}_\vp(t) \ {\rm e}^{i\Theta(\vp,t)}\,\,\bigg]\, \,,\\ 
{\dot   \chi}^{(+)}(\vp,t)&=&\quad i\omega(\vp,t)\ \bigg[  \alpha^*_\vp(t) \
\phi_\vp^{(+)}(t) \ {\rm e}^{i\Theta(\vp,t)}- \beta_\vp(t) \
\phi^{(-)}_\vp (t) \ {\rm e}^{-i\Theta(\vp,t)}\bigg]\,\,.
\eea
These new conditions together with the ansatz
\be\label{210}
\phi^{(\pm)}_\vp(t)=\sqrt{\frac{\omega(\vp,t)\pm P_\parallel(t)}{\omega(\vp,t)}} \ ,
\ee 
enable us to extract the  differential equation for the Bogoliubov coefficients 
\bea
\label{240}  
{\dot \alpha}_\vp(t)&=&\quad {\ds\frac{eE(t)\varepsilon_\perp}{2\omega^2(\vp,t)}}\
\beta^*_\vp(t) \ {\rm e}^{2i\Theta(\vp,t)}\,\,,\\ 
{\dot \beta}^*_\vp(t)&=&-{\ds\frac{eE(t)\varepsilon_\perp}{2\omega^2(\vp,t)}}\
\alpha_\vp(t) \ {\rm e}^{-2i\Theta(\vp,t)}\,\,.
\eea
by differentiating (\ref{190}) and using (\ref{40}) and (\ref{170}).

After integration of above equations we get the new  coefficients describing the instantaneous state at the time $t$.
\bea
\label{220}
\alpha_\vp(t)&=&{\ds\frac{1}{2\sqrt{\omega(\vp,t) \
(\omega(\vp,t)-P_\parallel(t))}}}\bigg(\omega(\vp,t)\ \chi^{(-)}(\vp,t)+i \ {\dot
\chi}^{(-)}(\vp,t)\bigg) \ {\rm e}^{i\Theta(\vp,t)}\,\,,\\ 
\beta^*_\vp(t)&=&-{\ds\frac{1}{2\sqrt{\omega(\vp,t) \
(\omega(\vp,t)-P_\parallel(t))}}}\bigg(\omega(\vp,t) \ \chi^{(-)}(\vp,t)-i \ {\dot
 \chi}^{(-)}(\vp,t)\bigg) \ {\rm e}^{-i\Theta(\vp,t)}\,\,.
\eea

It is convenient to redefine operators in order to absorb the dynamical phase
\bea
\label{260} 
B_{\vp r}(t)&=&b_{\vp r}(t) \  e^{-i\Theta(\vp,t)}\,,\\
D_{\vp r}(t)&=& d_{\vp r}(t)\  e^ {-i\Theta(\vp,t)} 
\eea
while preserving  the anti-commutation relations:
\be 
\{B_{\vp r}(t),B^+_{\vp ' r'}(t)\}=\{D_{\vp r}(t),D^+_{\vp ' r'}(t)\}= \delta_{rr'} \ \delta_{\vp \vp '} \, . 
\ee
It is straightforward to show  that these redefined operators satisfy the Heisenberg-like  equations
of motion 
\bea
\label{270} 
\ds\frac{dB_{\vp r}(t)}{dt}&=&-\frac{e E(t)\varepsilon_\perp}{2\omega^2(\vp,t)} \ D^+_{-\vp r}(t) 
+ i \ [H(t), \ B_{\vp r}(t)]\,\,, \\ 
\ds\frac{dD_{\vp r}(t)}{dt}&=&\phantom{-}\frac{e E(t)\varepsilon_\perp} {2\omega^2(\vp,t)}\ B^+_{-\vp r}(t)
+i \ [H(t),\ D_{\vp r}(t)] \ ,
\label{271} 
\eea
where $H(t)$ is the hamiltonian of the quasiparticle system
\be
\label{280}
H(t)=\sum_{r,\vp} \omega(\vp,t)\bigg(B^+_{\vp r}(t) \
B_{\vp r}(t)-D_{-\vp r}(t) \ D^+_{-\vp r}(t)\bigg)\,\, .
\ee
The first term on the r.h.s. of Eqs.~(\ref{270}) and (\ref{271}) is caused by
the unitary non-equivalence of the in-representation and the
quasiparticle one.

On this stage of the calculations  we can construct the distribution
function of electrons (with the momentum $\vp$ and  spin $r$) 
\be
\label{300}
f_r(\vp,t) = \langle 0_{\rm in}|b^+_{\vp r}(t) \ b_{\vp r}(t)|0_{\rm in}\rangle
= \langle 0_{\rm in}|B^+_{\vp r}(t) \ B_{\vp r}(t)|0_{\rm in}\rangle \,\,
\ee
and positrons
\be \label{320}
{\bar f}_r(\vp,t) = \,\langle 0_{\rm in}|d^+_{-\vp r}(t) \ d_{-\vp r}(t)|0_{\rm in}\rangle \,
=\,\langle 0_{\rm in}|D^+_{-\vp r}(t) \ D_{-\vp r}(t)|0_{\rm in}\rangle \,\,.
\ee

Charge conservation implies  $f_r(\vp,t) = {\bar f}_r(\vp,t)$, so that
summation over momentum and spin gives the normalization to the total number of pairs at a given 
time $t$
\be
\label{330} 
\sum \limits_{r,\vp}f_r(\vp ,t)= \sum \limits_{r,\vp} \bar f_r(\vp ,t)=N(t)\ . 
\ee

Now the differentiation of Eq.~(\ref{300}) w.r.t. time and the use of  the equation of motion (\ref{270}) 
results in 
\be
\label{340}
\frac{d f_r(\vp,t)}{dt}= -\frac{eE(t)\varepsilon_\perp}{\omega^2(\vp,t)} \
{\rm Re}\{\Phi_r(\vp,t)\} \,\,,
\ee
where the function 
\be 
\label{350}
\Phi_r(\vp,t)= \langle 0_{\rm in}|D_{-\vp r}(t) \ B_{\vp r}(t)|0_{\rm in}\rangle \,\,
\ee
governs the reaction of the QED vacuum in the presence of an external electric field. 
\newline
Differentiation of (\ref{350}) requires again the use of the equation of motion (\ref{270}) and together with $f_r(\vp,t) = {\bar f}_r(\vp,t)$ leads to

\be
\label{360}
\frac{d\Phi_r(\vp,t)}{dt}=\frac{e E(t)\varepsilon_\perp}{2\omega^2(\vp,t)}\bigg[2f_r(\vp,t)-1\bigg]-2i\omega(\vp,t) \
\Phi_r (\vp,t)\,\,.
\ee
The solution of Eq.~(\ref{360}) has the  integral form
\be
\label{370}
\Phi_r(\vp,t) = \frac{\varepsilon_\perp}{2}\int_{t_0}^t dt'\frac{e E(t')}{\omega^2(\vp,t')}\bigg[2f_r(\vp,t')-1\bigg]{\rm e}^{2i[\Theta(\vp,t')-\Theta(\vp,t)]}\,\,.
\ee
It is straightforward to see that $\Phi_r(\vp,t)\big|_{t = t_0}$ vanishes when $A(t_0) =0 $.

Now we are ready to write the expression (\ref{340}) in the full form
\be
\label{380} 
\frac{df_r(\vp,t)}{dt}= \frac{e E(t)\varepsilon_\perp}{2\omega^2(\vp,t)}\int_{t_0}^t dt' 
\frac{e E(t')\varepsilon_\perp}{\omega^2(\vp,t')}\bigg[1-2f_r(\vp,t')\bigg]
\cos \bigg(2[\Theta(\vp,t)-\Theta(\vp,t')]\bigg)\,\,,
\ee
which is the wanted kinetic equation for particle production.
Since the distribution function  does not depend on spin (\ref{380}), we can
skip the index $r$.

\subsection{Creation of boson pairs}

The derivation procedure of the kinetic equation for bosons is similar to that of fermions.
The starting point is the Klein-Gordon equation
\be
\label{kg}
\bigg((\partial^\mu + ieA^\mu)(\partial_\mu+ieA_\mu)+m^2\bigg)\phi(x)=0\,
\ee
with the external vector potential $A_\mu =(0,0,0, A(t))$.
In this case the ansatz for the solution is given in the form \cite{Bjorken:1965zz}
\be
\label{A1}
\phi^{(\pm)}_{\vp}(x)=[2\omega(p)]^{-1/2}\ e^{i\bar x\vp} g^{(\pm)}(\vp ,t)\ ,  
\ee 
where the functions $g^{(\pm)}(\vp ,t)$ are analogues of ${\chi}^{(\pm)}(\vp,t)$ from (\ref{40}) and
satisfy the oscillator-type equation 
\be
\label{A2} 
\ddot g^{(\pm)}(\vp ,t)+\omega^2 (\vp,t) \ g^{(\pm)}(\vp ,t)=0\ .
\ee
The field operator in the in-state is defined as 
\be
\label{A4}
\phi (x)=\int d^3p \ [\ \phi^{(-)}_{\vp}(x) \ a_{\vp}(t_0)+
\phi^{(+)}_{\vp}(x) \ b^+_{\-vp}(t_0)\ ]\,.
\ee
The diagonalization of the Hamiltonian with respect  to the instantaneous states is achieved by the transition to the quasiparticle representation.  
The Bogoliubov transformation for creation and annihilation operators of quasiparticles has the form
\bea
\label{A5}
a_{\vp }(t) &=&\alpha_\vp (t)\ a_{\vp }(t_0) + \beta_\vp (t)\ b^+_{-\vp }(t_0)\ ,\\ 
b_{-\vp }(t)&=&\alpha_{-\vp}(t)\ b_{\vp }(t_0)+\beta_{-\vp}(t)\ a^+_{-\vp }(t_0)\  
\eea
with the condition 
\be
\label{A6}
|\alpha_\vp(t)|^2-|\beta_\vp(t)|^2=1\,\,.  
\ee
The calculation of the coefficients $\alpha$ and $\beta$ requires similar steps like in the fermion case.
We obtain the equations of motion for the coefficients of the canonical transformation (\ref{A5}) as follows
\bea
\label{A12}
\dot\alpha_\vp (t)&=&\frac{\dot\omega(\vp ,t)}{2\omega(\vp ,t)} \ \beta^*_\vp(t) \ e^{2i\Theta (\vp ,t)}\,,\\
 \dot\beta_\vp(t)&=& \frac{\dot\omega(\vp ,t)}{2\omega(\vp ,t)}\ \alpha^*_\vp (t) \ e^{2i\Theta (\vp ,t)}\,. 
\eea
Following the derivation procedure for the case of fermion production,  we get
\be
\label{380b} 
\frac{df_r(\vp,t)}{dt}= \frac{e E(t)p_\parallel}{2\omega^2(\vp,t)}\int_{-\infty}^t dt' 
\frac{e E(t')p_\parallel}{\omega^2(\vp,t')}\bigg[1+2f_r(\vp,t')\bigg]\cos \bigg(2[\Theta(\vp,t)-\Theta(\vp,t')]\bigg)\,\,.
\ee

\section{Discussion of the source term}
\subsection{Properties of the source term}
In the previous section the derivation of the kinetic equation has been given. 
We can give a combined result for the Schwinger source term for bosons(+) and fermions(-)
on the right hand side of the kinetic equation  
\be
\label{380fb} 
S_{\pm}=
\frac{1}{2}\lambda_{\pm}(\vp,t)\int_{t_0}^t dt' \lambda_{\pm}(\vp,t')\bigg[1\pm 2f(\vp,t')\bigg]\cos
\bigg(2[\Theta(\vp,t)-\Theta(\vp,t')]\bigg)\,\,,
\ee
where
\begin{eqnarray}
\lambda_{-}(\vp,t) = e E(t)\varepsilon_{\bot}/\omega^{2}(\vp,t)
\, ,\qquad   \lambda_{+}(\vp,t) 
= eE(t)\mbox{p}_\parallel/\omega^{2}(\vp,t) \label{lambda}.
\end{eqnarray} 
This source term does not include the effects of back reaction of created pairs on the electric field and $e^{+}e^{-}$ collisions. 
Nevertheless it possesses interesting properties which we enumerate in the following. 
\begin{enumerate}
\item{
The source term is of non-Markovian type. 
This means that memory effects are present in the pair creation process. 
In our case the non-Markovianity  comes in via the statistical factor $1\pm 2f(\vp,t)$ under time integral in (\ref{380fb}), which makes the term dependent on the whole pre-history of $f(\vp,t)$.
Studies done by Rau \cite{Rau:1994ee} also have shown the non-Markovian character of the particle production process. 
They have used a projection method so one can conclude that the memory effects in $S_{\pm}$ are not artefacts.}

\item{ The source term is characterized by three time scales: the time scale of the external field, the memory time 
\be
\tau_{\rm mem}\sim \frac{\varepsilon_\perp}{eE}
\ee
and the production interval
\be
\tau_{\rm prod}=1/<S_{\pm}>\,,
\ee
where  $\langle S_{\pm} \rangle$ denotes the time averaged production rate.
The Markovian limit (approximate absent of statistical  factor) is reached  when  
$\tau_{\rm mem} \ll \tau_{\rm prod}$. 
This condition translates to  $E \ll m^2/e< \varepsilon^2_\perp/e$. 
For the low density limit and constant field the equation (\ref{380b})  resemble results  of 
Rau \cite{Rau:1994ee}.}

\item{The difference of the dynamical phases,
$ \Theta(\vp ,t)-\Theta(\vp ,t') $, under the integral in $S_{\pm}$ is the source of the high frequency oscillations related to Zitterbewegung.  }

\item{ 
The source term of the form (\ref{380fb}) causes entropy production (see also \cite{Rau:1994ee}) and thus is the reason of time irreversibility. 
However, the increase is non-monotonic due to lack of collision \cite{Habib:1995ee,Smolyansky_2012}. 
}

\item{
Particles are produced with non-zero momentum in contrast to previous studies, e.g. Ref.~\cite{Kluger:1992md}.}

\item{
In the case of the low-density limit and a constant electric field one can reproduce Schwinger's formula and Rau's results \cite{Rau:1994ee}) 
\be\label{430} {\cal S}^{{\rm cl}}=\lim_{t\to +\infty}(2\pi)^{-3}g\int d^3P \ {\cal S}(\wvp,t)=\frac{e^2E^2}{4\pi^3}\exp \bigg(-\frac{\pi m^2}{|eE|}\bigg) \ .
\ee}
\end{enumerate}

In the next subsection we discuss the Markovian and the low-density limits more in detail.

\subsection{Low-density  and Markovian limit}

In this section we discuss in short two related approximations which are applicable when the external electric field is considerably smaller than the critical field $E \ll E_c$.
These are the Markovian approximation and the low-density limit.

The Markovian limit of the non-Markovian source term (\ref{380fb}) is defined by the neglect of memory effects, i.e. by replacing the argument of the distribution function in the source term $f(\vp,t')\to f(\vp,t)$
so that it becomes independent of the prehistory of the distribution function  and the KE takes the form
\be
\label{40}
 \frac{d\,f^M_\pm(t)}{dt}= [1\pm
2f^M_\pm(t)]S^0_\pm(t)=S^M_\pm(t),   
\ee
where
\be
\label{380lowdensity} 
S^{0}_{\pm}(t)=
\frac{1}{2}\lambda_{\pm}(\vp,t)\int_{t_0}^t dt' \lambda_{\pm}(\vp,t')\cos
\bigg(2[\Theta(\vp,t)-\Theta(\vp,t')]\bigg)\,\,.
\ee
is the source term in the low-density limit when $f(\vp,t) \ll 1$ for every time $t$ so that the statistical
factor becomes trivial $1\pm 2f(\vp,t) \approx 1$.

Together with the initial condition  $f(t_0) = 0$ the Markovian KE (\ref{40}) has the solution 
\be
\label{50}
f^M_\pm(t)=\mp\frac{1}{2} \bigg(1-\exp\bigg[\pm 2\int_{t_0}^t d\,t'S^0_\pm(t')\bigg]\bigg)\,.
\ee
The lowest order expansion of Eq. (\ref{50}) w.r.t. the source term results in the low density solution
\be
\label{55}
f^0_\pm(t) = \int_{t_0}^t d\,t'S^0_\pm(t')\,.
\ee
Both low density approximation and Markovian limit hold only when  $E \ll E_c$.

The low density limit gives us a tool to prove the positive definiteness of the distribution function. 
Using the trigonometric identity 
$\cos(\alpha -\beta)= \cos\alpha\cdot\cos\beta + \sin\alpha\cdot\sin\beta$ we rewrite (\ref{55}) as 
\bea
\label{60}
f^0_\pm(t)=\frac{1}{2} \int\limits ^t_{t_0}d\,t'g^1_\pm(t')
\int\limits ^{t'}_{t_0}d\,t''g^1_\pm(t'')
+ \frac{1}{2} \int\limits
^t_{t_0}d\,t' g^2_\pm(t')\int\limits ^{t'}_{t_0}d\,t''g^2_\pm(t'')\,,
\eea
\be\nonumber
g^{1,2}_\pm(\tau)={\lambda}_\pm(\tau)
\left\{\begin{array}{c}\cos[2\Theta (\tau)]\\ \sin[2\Theta
(\tau)]\end{array}\right\}.
\ee
The next step is done with the standard trick 
\begin{equation}
\label{trickDysonSeries}
 \int\limits ^t_{t_0}d\,t'A(t')
\int\limits ^{t'}_{t_0}d\,t''B(t'') = \frac{1}{2}\int\limits ^t_{t_0}d\,t'A(t')
\int\limits ^{t}_{t_0}d\,t''B(t'')
\end{equation} 
used in the  derivation of the Dyson series. 
The application of (\ref{trickDysonSeries}) to (\ref{60}) leads to the quadratic form 
 \be
 \label{70}
 f^0_\pm(t)=\frac{1}{4} \bigg(\int\limits
 ^t_{-\infty}d\,t'g^1_\pm(t')\bigg)^2 + \frac{1}{4} \bigg(\int\limits
  ^t_{-\infty}d\,t'g^2_\pm(t')\bigg)^2  \ . 
 \ee
 Now it straightforward to see that the distribution function is positive definit as it is required by the kinetic theory
 \be
 f^0_\pm(t)\ge 0 \ .
 \ee
 More detailed studies of the kinetic equation and its limiting cases have been presented in 
 \cite{Schmidt:1998zh}.
 
 \subsection{Entirely differential form of the kinetic equation }

The numerical solution of the integro-differential equation
\be
\label{integrodiff} 
\frac{df_r(\vp,t)}{dt}=
\frac{1}{2}\lambda_{\pm}(\vp,t)\int_{t_0}^t dt' \lambda_{\pm}(\vp,t')\bigg[1\pm 2f(\vp,t')\bigg]
\cos \bigg(2[\Theta(\vp,t)-\Theta(\vp,t')]\bigg)\,\,
\ee
although being straightforward is highly ineffective due to the double time integration.
First of all one need to deal with the fastly oscillating term  $\cos(2[\Theta(\vp,t)-\Theta(\vp,t')]) $. 
To address this problem we can make the integration step small enough. 
However, due to the non-Markovian character of the equation the storage of entire prehistory of $f(t)$  in the computer memory is required. 
Luckily one can avoid these complications by transforming (\ref{integrodiff}) to the equivalent system of
time local ordinary differential equations \cite{Bloch:1999eu}.    
In order to perform the transformation we introduce two auxiliary functions
\bea
\label{auxV} 
v(t)&=&\int_{t_0}^t dt' \lambda_{\pm}(\vp,t')\bigg[1\pm 2f(\vp,t')\bigg]
\sin \bigg(2[\Theta(\vp,t)-\Theta(\vp,t')]\bigg)\,\,,
\\
\label{auxU} 
u(t)&=&\int_{t_0}^t dt' \lambda_{\pm}(\vp,t')\bigg[1\pm 2f(\vp,t')\bigg]
\cos\bigg(2[\Theta(\vp,t)-\Theta(\vp,t')]\bigg)\,\,.
\eea
The differentiation of these functions with respect to $t$ together with
\begin{equation}
\partial_t \bigg(2[\Theta(\vp,t)-\Theta(\vp,t')]\bigg) = \partial_t  \int_{t'}^t \omega(\vp,t'') dt'' = \omega(\vp,t)
\end{equation}
 yields
\bea
\dot{v} &=& \lambda_{\pm}(\vp,t)\bigg[1\pm 2f(\vp,t)\bigg] - 2\omega(\vp,t)u(t) ,
\\
\dot{u}&=& 2\omega(\vp,t)v(t)~.
\eea
Taking into account (\ref{integrodiff}) results in the system of first order coupled  differential equations 
\begin{eqnarray}
\label{ode}
\dot{f} &= &\lambda_{\pm}(\vp,t)v(t) ,
\\
\dot{v} &=& \lambda_{\pm}(\vp,t)\bigg[1\pm 2f(\vp,t)\bigg]
- 2\omega(\vp,t)u(t) ,
\\
\dot{u}&=& 2\omega(\vp,t)v(t)
\end{eqnarray}
with the initial condition 
\begin{equation}
f(t_0) = u(t_0) = v(t_0) = 0~.
\end{equation}
The above system is much simpler to solve numerically. 

\section{Vacuum particle-antiparticle creation in strong fields as a field induced phase transition}

The functions $u(\mathbf{p},t)$ and $v(\mathbf{p},t)$ are defined by the anomalous averages
\begin{eqnarray}
\label{ano_ave}
    f^{(+)}(\mathbf{p},t)&=&\langle \mathrm{in}|B^{+}(\mathbf{p},t)D^{+}(-\mathbf{p},t)|\mathrm{in}\rangle, \\
    f^{(-)}(\mathbf{p},t)&=&\langle\mathrm{in}|D(-\mathbf{p},t)B(\mathbf{p},t)|\mathrm{in}\rangle \label{ano_ave2}
\end{eqnarray}
by means of the relations \cite{Schmidt:1998vi, Pervushin:2003uu}
\begin{eqnarray}\label{relations}
   u(\mathbf{p},t)&=&2\mathrm{Re}f^{(+)}(\mathbf{p},t)=2\mathrm{Re}f^{(-)}(\mathbf{p},t) \\
   v(\mathbf{p},t)&=&2\mathrm{Im}f^{(+)}(\mathbf{p},t)=-2\mathrm{Im}f^{(-)}(\mathbf{p},t) \, .
\end{eqnarray}
Hence their combination $\Phi(\mathbf{p},t)=u(\mathbf{p},t)+iv(\mathbf{p},t)$ fulfills the role of the order parameter for the phase transition from the vacuum where $\Phi(\mathbf{p},t)=0$ is set as initial condition, to $\Phi(\mathbf{p},t)\neq 0$ for the REPP.
The appearance of a nonvanishing value for this order parameter  is stipulated by the violation of the time inversion symmetry of the Hamiltonian as induced by the time dependent external field.
The creation and annihilation operators $a^{(\pm)}(\mathbf{p},t)$ and $b^{(\pm)}(\mathbf{p},t)$ of the electrons and positrons are used in Eqs.~(\ref{ano_ave}) and (\ref{ano_ave2}) in the quasiparticle representation, in which the Hamiltonian of the system is diagonal.

The order parameter $\Phi(\mathbf{p},t)$ obeys to the equation of motion
\begin{equation}
\label{e_m_f}
   \dot{\Phi} = \lambda(1-2f) + 2i\omega\Phi ,
\end{equation}
which follows from the system of equations (\ref{ode}).
For any finite field with $E(t) \to 0$ and $A(t) \to A_{\rm out}$ at $t \to \infty$ we have $\lambda (t) \to 0$ and 
\begin{equation}
\label{epsilon}
   \omega (t) \to \omega_{\rm out} = \sqrt{ \varepsilon^2_{\bot} + (p_\parallel -eA_{\rm out})^2 }.
\end{equation}
The order parameter in this asymptotics oscillates with the frequency $2\varepsilon_{\rm out}$:
\begin{equation}
\label{op_out}
   \Phi(t) \to \Phi_{\rm out}(t) \sim \exp(2i\omega_{\rm out}t) .
\end{equation}
Thus, $|\Phi_{\rm out}(t)|^2 = {\rm const}$ after switching off the external field, i.e. the long-range order is formed. Such situation is typical for a phase transitions in systems with broken symmetry.

In the low density approximation $2f \ll 1$, the KE (\ref{380}) has a closed formal solution in the form 
of a useful quadrature formula  \cite{Schmidt:1998zh}
\begin{equation}\label{ld}
f(\mathbf{p} ,t) = \frac{1}{2}\int\limits^t_{t_0} dt^{\prime} \lambda(\mathbf{p}
,t^{\prime} )\int\limits^{t^{\prime}}_{t_0} dt^{\prime \prime} \lambda(\mathbf{p} ,t^{\prime \prime})\cos\theta(t^{\prime},t^{\prime \prime})~.
\end{equation}

The total number density of pairs is defined as
\begin{equation}\label{dens}
    n(t) = 2 \int \frac{d\mathbf{p}}{(2 \pi)^3} f(\mathbf{p} ,t)~,
\end{equation}
where the factor 2 corresponds to the spin degree of freedom.

In these lectures we discuss the numerical solution of the KE  (\ref{380}) for two relevant models of the electric field
\begin{itemize}
\item[(i)] the Eckart-Sauter field with characteristic duration of action $T$ 
(single-sheeted field)
\begin{equation} 
\label{field2}
E(t) = E_0 \cosh^{-2}(t/T),  \,  A(t)= -{TE_0} \tanh(t/T) , ~{\rm and} 
\end{equation}
\item[(ii)] the Gaussian envelope model of the laser pulse 
(multi-sheeted field) \cite{Alkofer_2009}
\begin{eqnarray} 
\label{field3} 
E(t) & = & E_0  \cos{ (\omega t) }\ e^{-t^2/2\tau^2 }, \\
A(t) &=& -\sqrt{\frac{\pi}{8}} E_0\tau \exp{(-\sigma^2/2)}\;\text{erf}\left(\frac{t}{\sqrt{2}\tau} -i\frac{\sigma}{\sqrt{2}}\right) + c.c. , \nonumber
\end{eqnarray}
where $\sigma= \omega \tau$ is a dimensionless measure for the characteristic duration of the pulse $\tau$ 
connected with the number of periods of the carrier field. 
\end{itemize}
The Eckart-Sauter field (\ref{field2}) admits an exact solution 
\cite{Grib:1994,Nikishov:1969tt,Fedotov:2010ue} and is thus a benchmark case. 

In order to introduce the Keldysh parameter $\gamma = E_c \omega/E_0 m$ for the discussion of the field model (\ref{field2}) one can use the substitution $\omega \to 1/T$ and the definition of the critical value of the electric field $E_c = m^2/e$.
In the limiting case $\gamma \ll 1$ the tunneling mechanism 
(with participation of an infinite number of photons) 
dominates, whereas for $\gamma \gg 1$ pair creation is driven by the absorption of few photons.

The vacuum oscillations (Zitterbewegung) play a crucial role in the mechanism of  vacuum EPP creation. 
The usual energy of vacuum oscillations $\varepsilon_0 = \sqrt{m^2 + \mathbf{p}^2}$ is transformed here to the quasienergy (\ref{eq:energy})
in the presence  of the time dependent electric field. 
The memory effect (non-Markovian character of the KE), 
the fastly oscillating factor with the phase (\ref{160}) and the 
frequency $2\varepsilon$ (the dynamical energy gap) are the essential elements in the KE (\ref{380}).
This equation contains two characteristic time scales: 
a slow one associated with the time scale of the external field period, $2\pi/\omega$, 
and a fast one given by the Compton time $\tau_c = 2\pi/m$. 
These scales are usually vastly different, $\omega \ll m$. 
The coupling of the dynamics related to these two scales leads to a very complicated structure of the distribution function, both in the first stage (generation of the quasiparticle EPP (QEPP)) and in the final stage (formation of the residual EPP (REPP)) \cite{Blaschke:2013ip}.

\section{Field induced phase transition \label{sect:2}}
In the considered situation, the FIPT appears as rearrangement of the vacuum state under the action of a classical electromagnetic field.
It leads to the $t -$ noninvariant quasiparticle vacuum which corresponds to a non-stationary Hamiltonian of the system (the Coleman theorem \cite{Grib:1994, Coleman:1966}).
In this connection, the quasiparticle electron-positron pairs are the massive analog of the Goldstone bosons \cite{Grib:1994,Glashow:1967rx}.
Let us consider {features of} the FIPT.

\subsection{Transient stage}

\subsubsection{One-sheeted field model} 

\begin{figure}[!htb]
\includegraphics[width=0.5\textwidth]{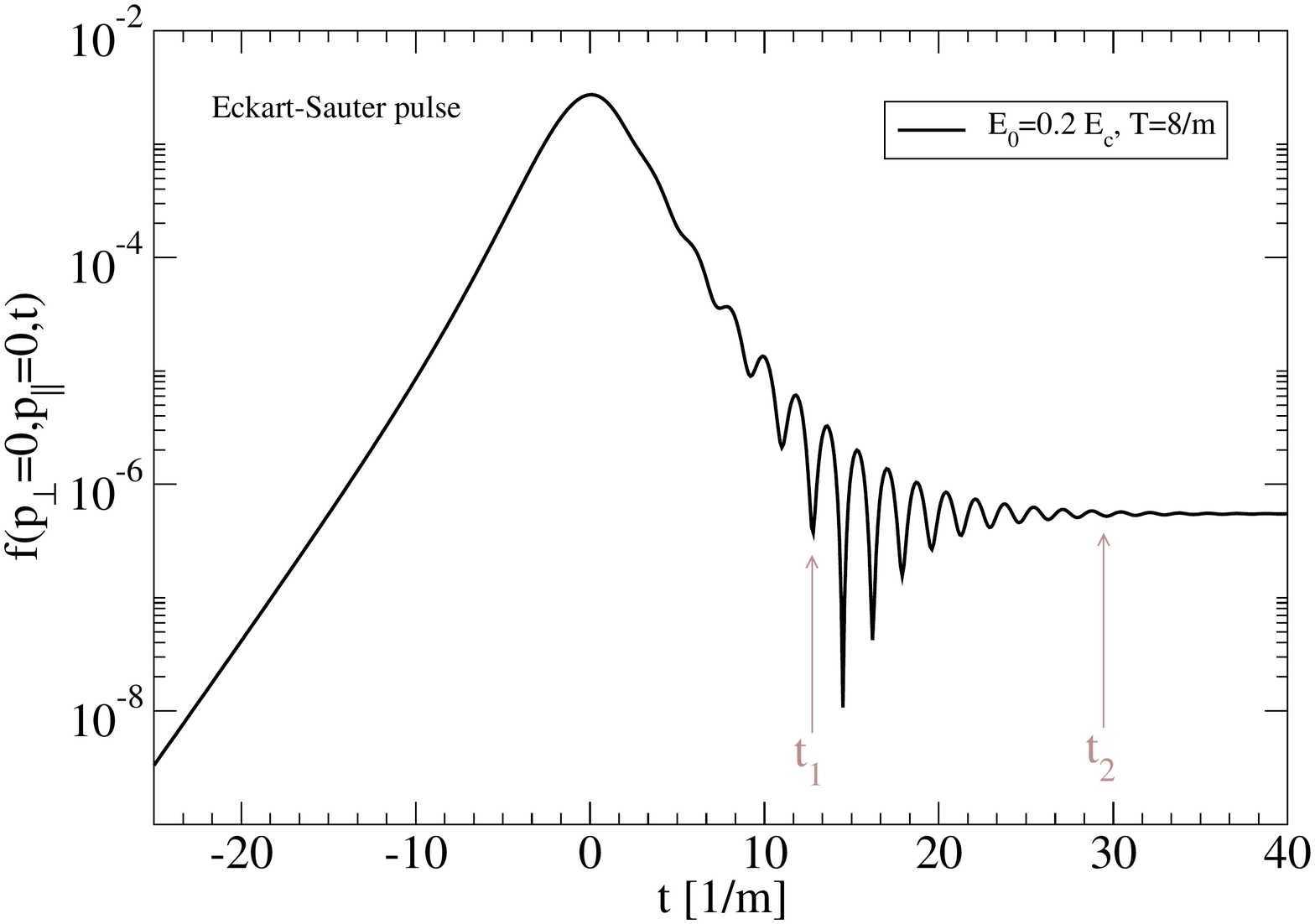} \hfill
\includegraphics[width=0.5\textwidth]{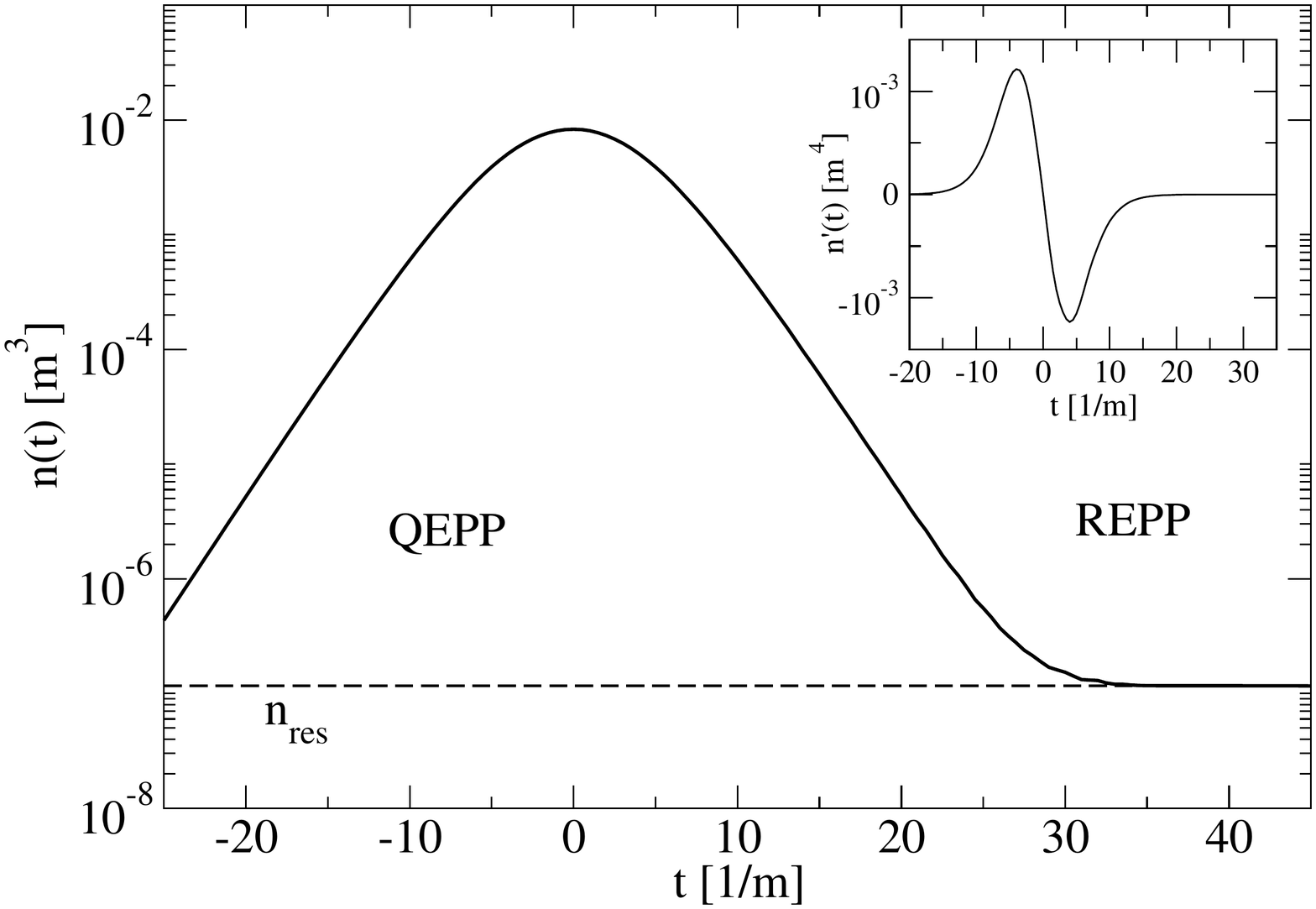}
\caption{The transition from the QEPP plasma to the final state for the Eckart-Sauter pulse type (\ref{field2}) 
with $E_0 = 0.2E_c$ and $T = 8$. 
The labels $t_{1}$ and $t_{2}$ denote approximately the begin and the end of the transient stage.
{\bf Left panel:} Evolution of the distribution function for the point $p_\bot =  p_\parallel = 0$.
{\bf Right panel:} Evolution of the pair number density (\ref{dens}) and the local pair production rate 
$w(t) = \dot n(t)$ (inset). 
\label{fig:1}}
\end{figure}

The typical picture of the EPP evolution under the action of the {one-sheeted} pulse (\ref{field2}) is presented in Fig.~\ref{fig:1}. 
The left panel shows that the transient process of the fast EPP oscillations divides the evolution of the EPP into two domains, the QEPP and the REPP. 
After momentum integration the fast oscillations of the transient process are smoothed out, see the right panel of Fig.~\ref{fig:1}. 
The inset of that panel shows the local production rate. 
The results of the numerical solutions of the KE (\ref{380}) (or (\ref{ode})) coincide with the exact solution \cite{Grib:1994,Nikishov:1969tt,Hebenstreit:2011pm}. 
On all figures the time and frequency are scaled with the electron mass.

For qualitative orientation one can introduce here the time interval of the strong oscillations limited by point $t_{1}$ of the begin (that can be identified with the moment when the oscillations of the distribution function reach for the first time the level of the REPP) and the end $t_{2}$ (corresponding to the moment when the mean level of oscillations approaches that of the REPP and the elongation of the oscillations is significantly reduced). 
This transient period of the Zitterbewegung separates the smoothed QEPP stage from the REPP stage.

\begin{figure}[!htb]
\includegraphics[width=0.5\textwidth]{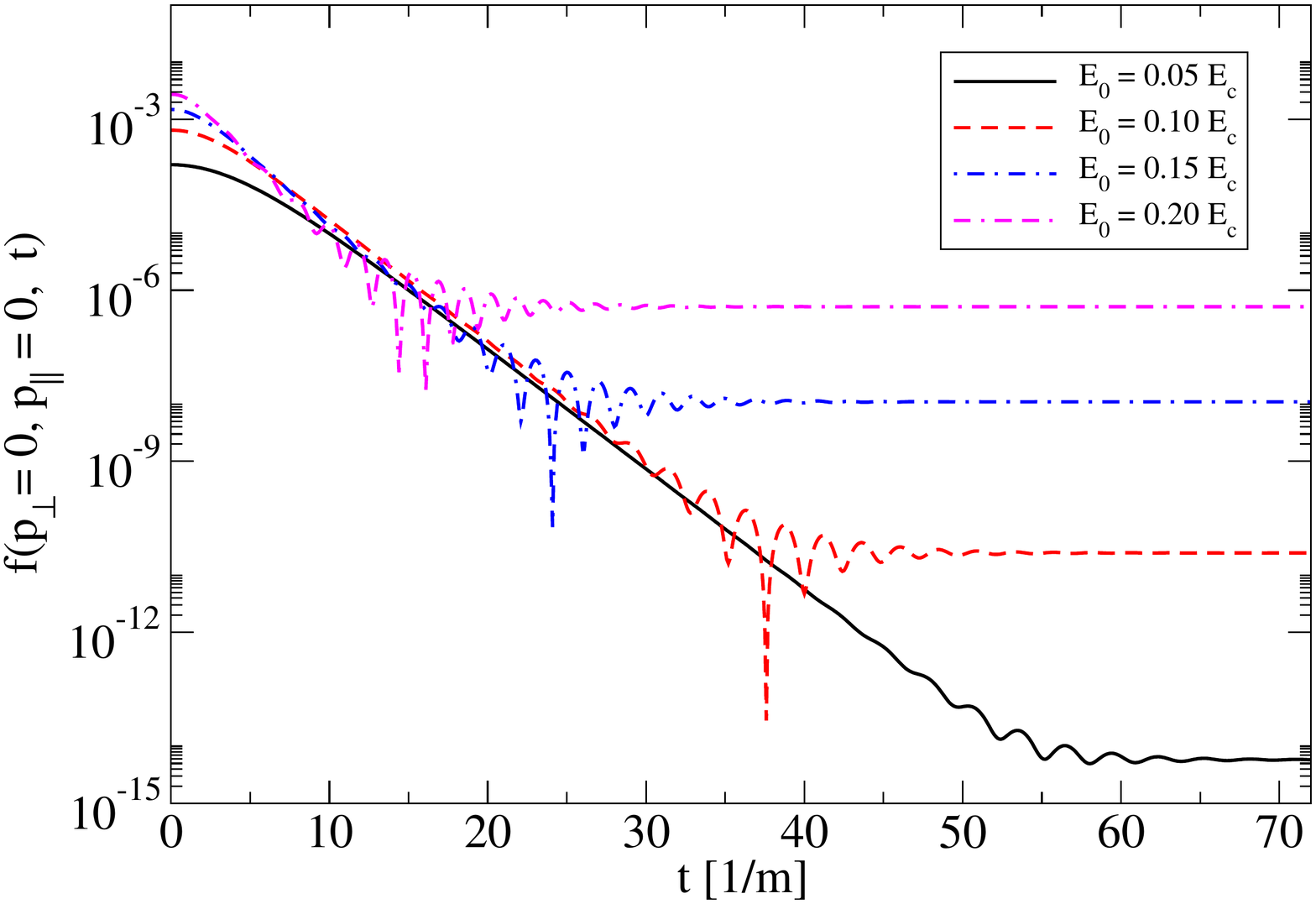} \hfill
\includegraphics[width=0.5\textwidth]{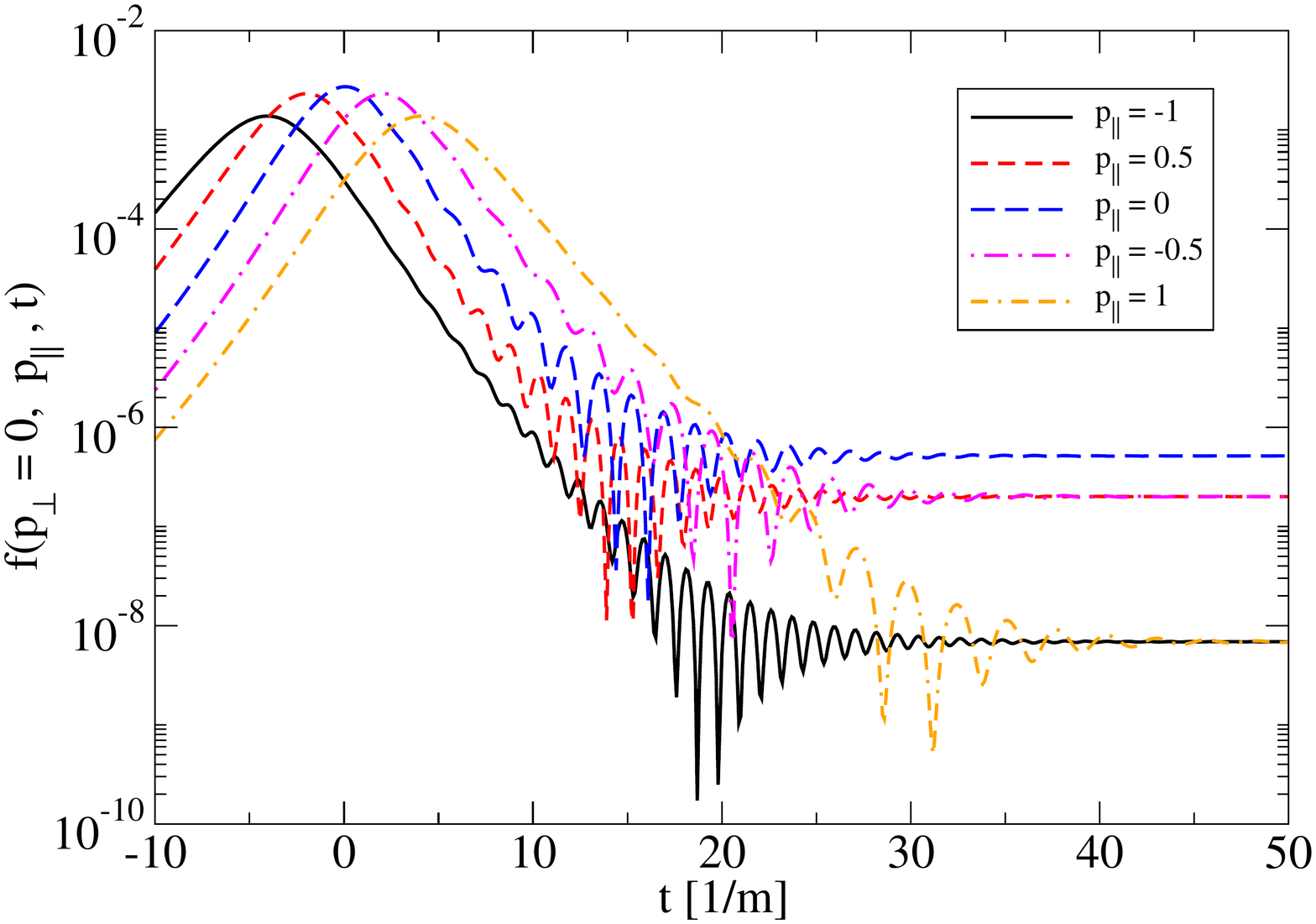}
\caption{Time evolution of the distribution function for the Eckart-Sauter pulse (\ref{field2}) with $T = 8.24$. 
{\bf Left panel:} At $p_\bot =  p_\parallel = 0$ for subcritical fields $E_0/E_c = 0.05, 0.10, 0.15$  and $0.20$. 
{\bf Right panel:} The details of phase transition at $E_0/E_c=0.2$ and $p_\bot =  0$ for different values  $p_\parallel = -1.0, -0.5, 0.0, 0.5, 1.0$.
\label{fig:2_}}
\end{figure}

Under similar conditions strong oscillations are observed also in other physical models with massive constituents. 
For example, they appeared in the domain of the relativistic phase transition with dynamical mass generation (the inertial mechanism of particle creation) including the Higgs mechanism 
\cite{Filatov:2007ha}. 
Their existence can be found also in the strong field dynamical models of strongly correlated systems 
(see, e.g., Ref.~ \cite{Oka:2011ct}).
Let us underline that the appearance of the transient region with strong oscillations takes place in the considered case of a smooth impulse  (\ref{field2}) without a carrier wave that would possess a high frequency component.

For a better understanding of this phenomenon let us consider the mechanisms of particle creation acting in the KE (\ref{380}) or in its approximate solution  (\ref{ld}). 
We will trace the evolution of the system in the smooth field (\ref{field2}) for $t>0$ which is accompanied by a field strength depletion. 
If the electric field is rather strong, for $t<t_{\rm 1}$ the acceleration mechanism represented by the force factor $eE(t)$ in the numerator of the amplitude (\ref{lambda}) is dominant whereas the fastly oscillating factor $\cos\theta(t,t^{\prime})$ on the r.h.s. of the KE (\ref{380}) smoothes out. 
The vicinity of the moment $t_{1}$ of the begin of the transient stage is characterized by a weakening of the accelerating field action and by the growth of the role of the fast oscillations with the frequency $2\varepsilon(\mathbf{p} ,t) \ge 2m$, in which one can neglect now the influence of a weak field so that the oscillation ``beard'' in the transient stage appears  Fig.~\ref{fig:1}.
The subsequent field depletion accompanied by the growth of the vector potential (and the quasi-momentum $P(t)$ (\ref{eq:p-long})) in the denominator of the amplitude (\ref{lambda}) leads to the asymptotic extinction of the oscillations and the approach of the final REPP state.

Fig.~\ref{fig:2_} demonstrates the fine structure of the distribution function in the transient period for varying field strength at fixed momentum (left panel) and for varying $p_\parallel = p_3$ at $p_\bot = 0$ and fixed field strength (right panel).
One can see from here that the behavior of the distribution function depends on the selection of a point 
$\mathbf{p}$ in momentum space. 
The maximum of the distribution function is realised for $p_\bot = 0$ in different points $p_\parallel = p_3$ at different time moments because the component $p_\parallel$ is contained in the amplitude (\ref{lambda}) together with the vector potential in the quasi-momentum $P(t)$ (\ref{eq:p-long}). 

The features of these oscillations are defined by the double quasienergy $2\varepsilon(\mathbf{p} ,t)$ and are reproduced well also by the numerical solution of  the KE. 

Transient area arises also in the case of an other exact solution with the infinite field model $A(t)= -E_0t$ \cite{Hebenstreit:2011pm}.

\subsubsection{Multi-sheeted field model} 

\begin{figure}[!thb]
\includegraphics[width=0.9\textwidth,height=0.6\textwidth]{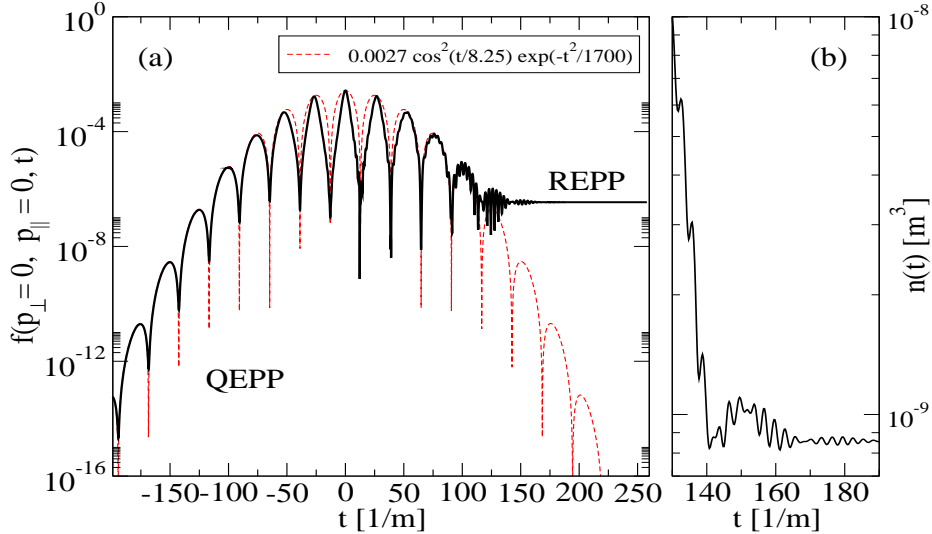}
\caption{Transition from QEPP to REPP in the case of a harmonic field with Gaussian envelope (\ref{field3}) with $E_0 = 0.2E_c$, $\sigma = 5.0$ and wavelength $0.02$ nm.
{\bf Left panel:} The distribution function for the point $p_\bot =  p_\parallel = 0$. 
The red dashed line shows the squared external electric field (\ref{field3}) for orientation.
{\bf Right panel:} The density $n(t)$ (\ref{dens}) in the region of the final transient bubble. 
\label{fig:3_}}
\end{figure}

\begin{figure}[!thb]
\includegraphics[width=\textwidth,height=0.7\textwidth]{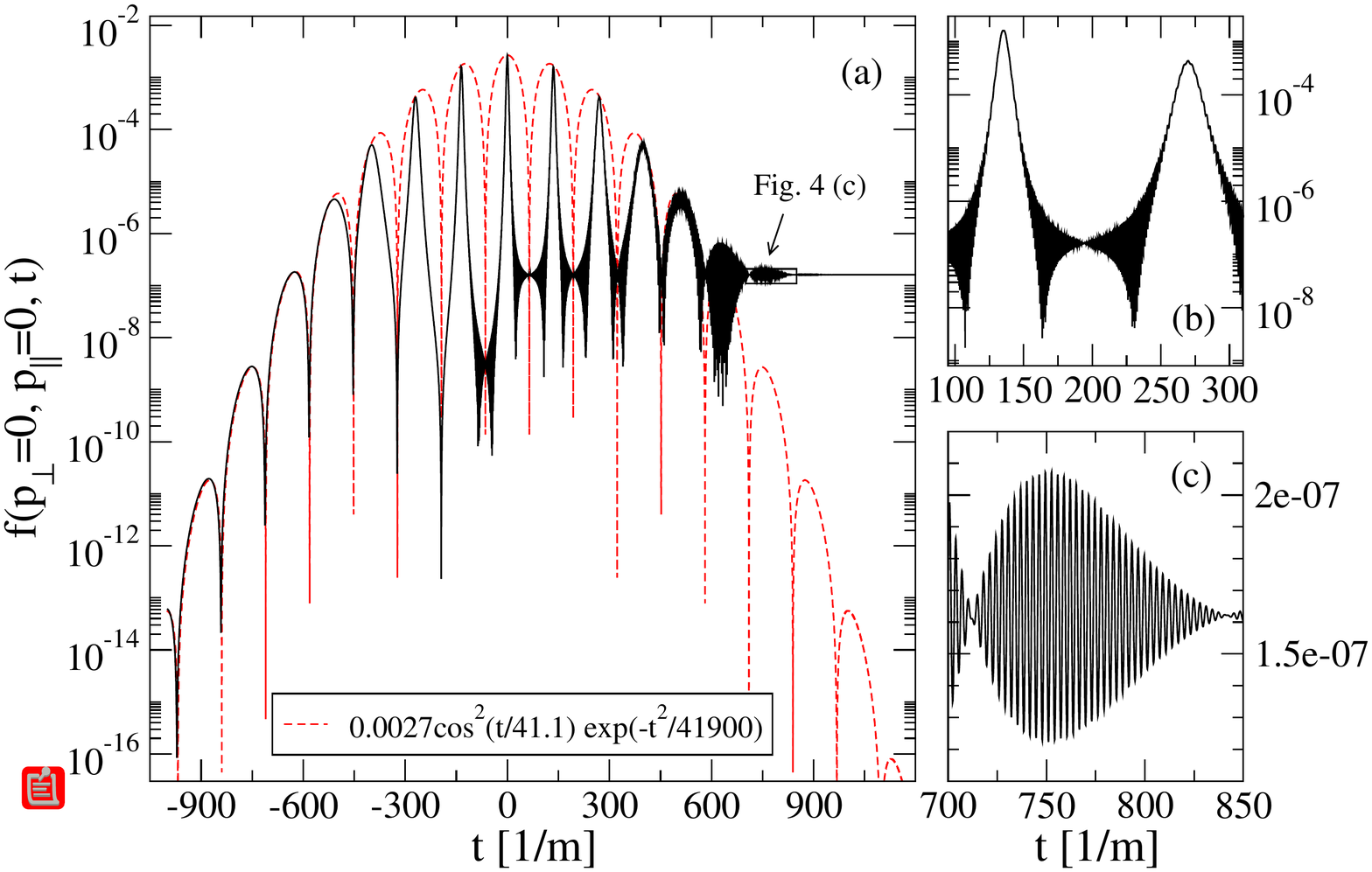}
\caption{Details of the evolution of the distribution function in the case of a harmonic field with the Gaussian envelope (\ref{field3}) with $E_0 = 0.2E_c$, $\sigma = 5.0$ and wavelength $0.1$ nm.
{\bf Left panel:} Evolution of the EPP over a large interval of 3000 Compton times.
The red dashed line shows the squared external electric field (\ref{field3}) for orientation.
{\bf Upper right panel:} Evolution of the EPP under the action of a single subpulse with a half-period duration. 
{\bf Lower right panel:}  The fine structure of the final transient bubble. 
\label{fig:5}}
\end{figure}

These features of the transient process are complicated in the case of a high frequency periodic field with a Gaussian envelope (\ref{field3}), $\omega \gg 1/\tau$ (multi-sheeted field model). 
Typical patterns are presented in Figs.~\ref{fig:3_} and \ref{fig:5}. 
In this case the neighboring intermediate smooth domains (peaks) separate the high-frequency transient stages. They are defined by the sub-pulses with a half-period $\pi/\omega$ of the external field (\ref{field3}).
The red dashed lines in Figs.~\ref{fig:3_} and \ref{fig:5} show the modulus of the external field (\ref{field3}) 
with the appropriately chosen normalization. 
The evolution is finished by the final intensive transient bubble, after which the EPP goes over to the out-state REPP.
The final transient bubble is traced also on the density curve of the EPP (the right panel of Fig.~\ref{fig:3_}. 
Apparently, this illustrates the effect of mutual amplification of EPP production as a result of the nonlinear interaction of the fast and slow components \cite{Dunne:2009gi,Otto:2015gla,Panferov:2015yda} of the field (\ref{field3}).

Fig.~\ref{fig:5}   demonstrates the evolution of the QEPP under the action of a periodical field  with  Gaussian  envelope  (19)  for another wavelength of the carrier signal, five times larger than the one in 
Fig.~\ref{fig:3_}.
This allows to show some details of the evolution more clearly.
Some features of this type of field appear in presence of repeated passages through the zero points of the 
field $E(t)$.
If such transitions occur sufficiently smoothly, then each of them has enough time to form a transition region, similar to the single "beard" in the left panel of Fig.~1.
The reduction of the carrier frequency allows to demonstrate in quite clear detail the repeated transient areas (see the upper right panel of Fig.~\ref{fig:5}).
With the decrease in the pulse amplitude they cease to form a stable QEPP (see the lower right panel of
Fig.~\ref{fig:5}).
As a result, we are seeing a stretched area of the transition to the REPP.
In this transition region vacuum fluctuations modulated acting field.
When turning off the field by the Gaussian envelope the modulation disappears
and the transition to the REPP state is completed.

The presence of a transient region of fast oscillations in the distribution function is characteristic for every field model. 
In this regard the discussed phase transition under the action of a strong electric field is a universal effect for quantum field systems with an energy gap.
We remark that in the case of massless 2+1 dimensional QED (e.g., for graphene), the high-frequency transient region is absent and the evolution of the particle-antiparticle plasma distribution function is smooth \cite{Panferov:2017}.

\begin{figure*}
\includegraphics[width=0.48\textwidth]{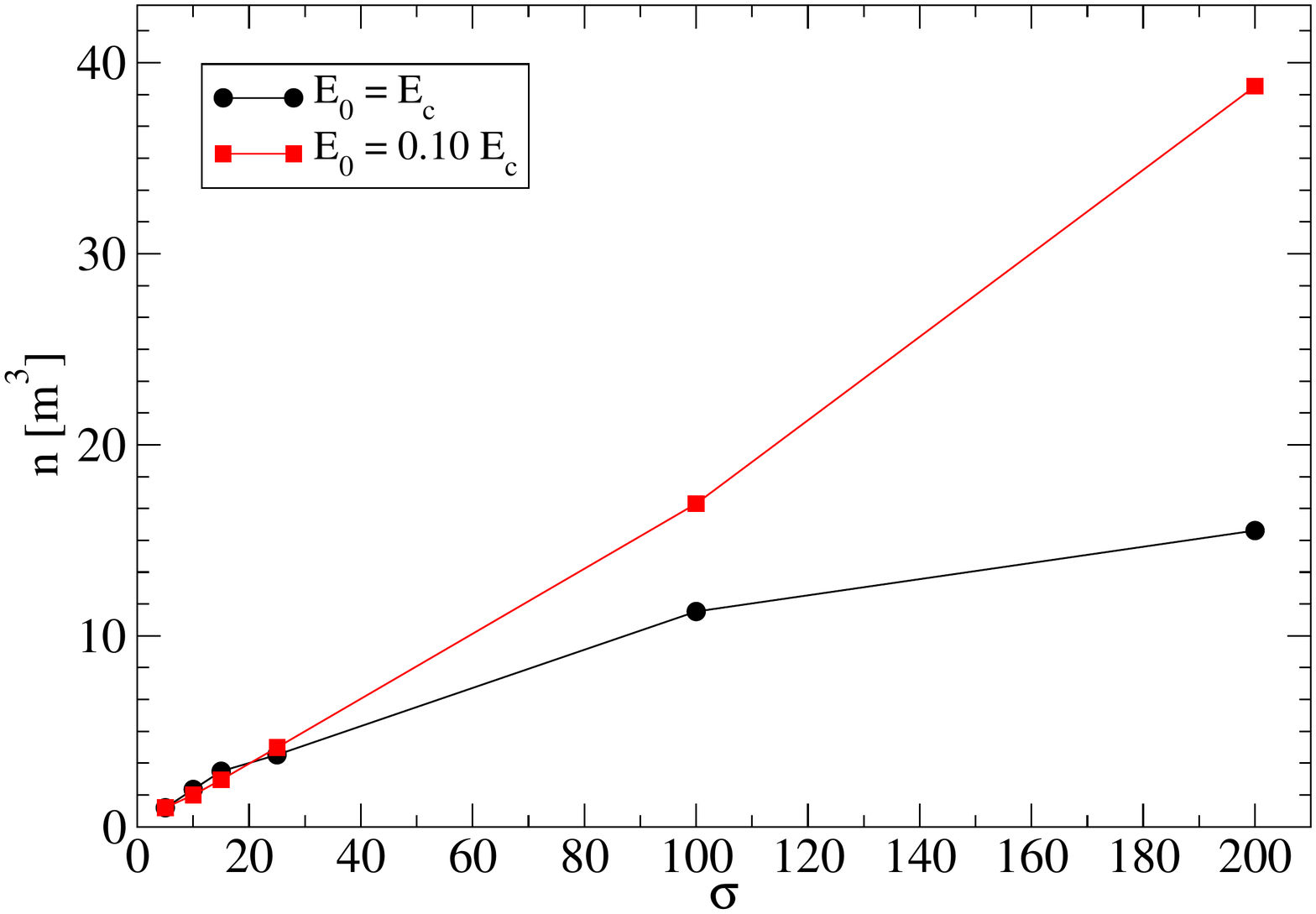} \hfill
\includegraphics[width=0.48\textwidth]{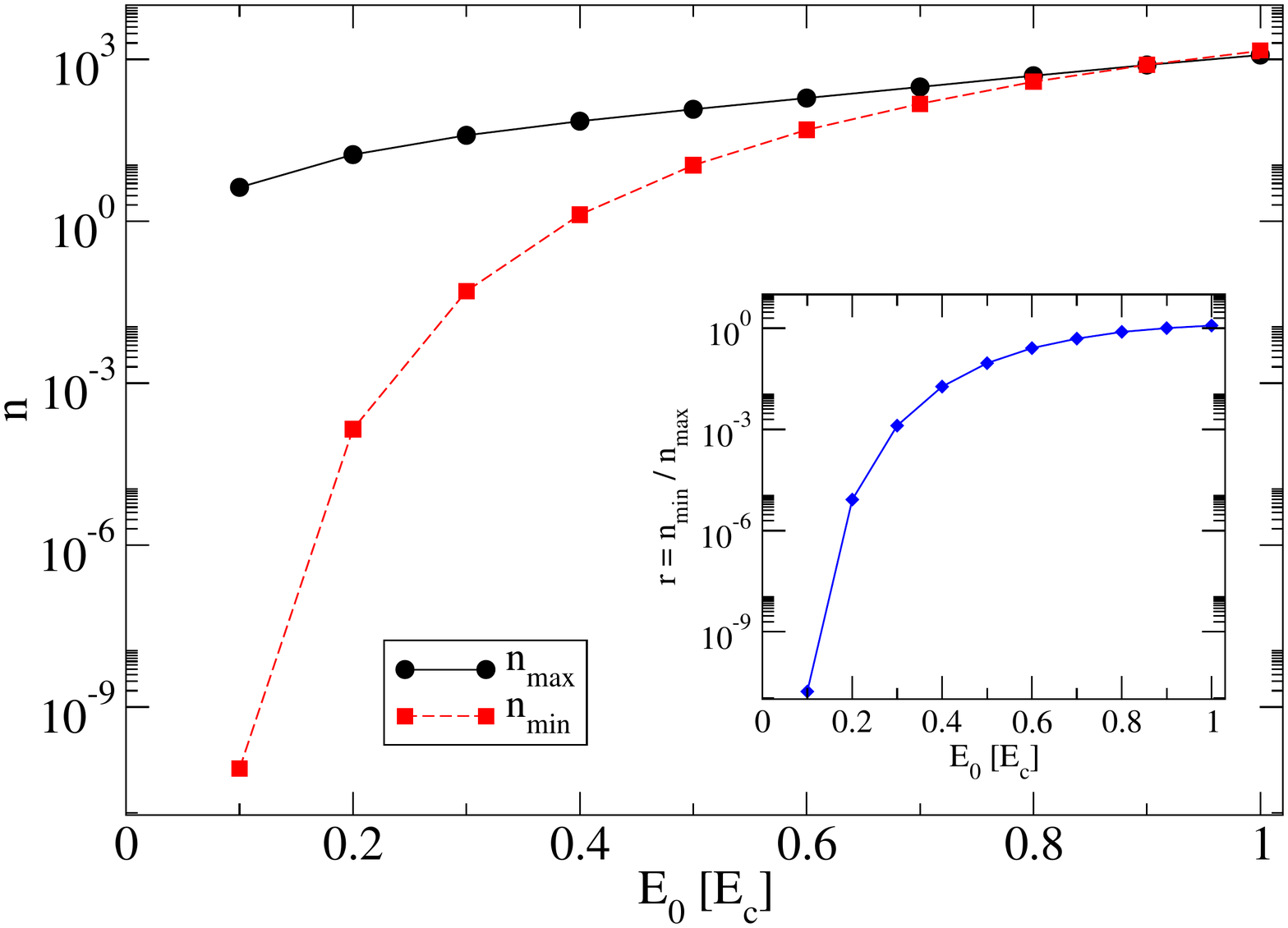}
\caption{Influence of the parameters the field pulse on REPP production.
{\bf Left panel:}  Influence of the laser pulse (\ref{field3}) duration $\sigma$ on the REPP density for the medium ($E_0/E_c=0.1$) and high ($E_0/E_c=1.0$) amplitude of the electric field. For both values of the field the unit value of density is determined as the density of the REPP produced by a pulse with a duration $\sigma = 5$.
{\bf Right panel:} Comparison of the maximum density of QEPP and density of the REPP in the range of values $0.1E_c \le E_0 \le 1.0E_c$ for the laser pulse (\ref{field3}) with a wavelength $0.02426~$nm 
($\omega = 0.1$) and $\sigma = 5$. The insertion shows the transformation coefficient 
$r(E_0/E_c) =n_{\rm res}/ n_{\rm max}$ for the shown values of the densities.
\label{fig:7}}
\end{figure*}

The left panel of Fig.~\ref{fig:7} demonstrates the dependence of the EPP pair density (\ref{dens}) on the pulse duration at fixed frequency $\omega$ in the field model (\ref{field3}). 
It exhibits a nonlinear accumulation effect for which the slope is approximately constant for weak fields 
whereas for strong fields we observe a saturation effect. 
Finally, the right panel of Fig.~\ref{fig:7} shows the dependence of the EPP pair density in the out-state in comparison to the maximal value attained within the entire period of the EPP evolution (see, e.g., Fig.~\ref{fig:1}, right).

\subsection{Strong nonequilibrium}

The entire process of vacuum EPP creation is a strong nonequilibrium one, including the final out-state.
In the first place, this conclusion follows from the exactly solvable models.
The distribution functions  of the out-state turn out to be the same for both, the constant field model $E(t) = E_0$ \cite{Hebenstreit:2011pm,Cooper:1992hw} and the Eckart-Sauter model (\ref{field2}) for $T \to \infty$ \cite{Gavrilov:1996pz, Grib:1994}
\begin{equation}
\label{f_degen}
f_{\rm out}(\mathbf{p}) = \exp \left[-\pi \frac{E_c}{E_0} \left(\frac{\varepsilon_{\bot}}{m}\right)^2 \right].
\end{equation}
This function is degenerate {with respect to} $p^3 = p_\parallel$ and therefore non-normalizable.
This leads to the necessity to extend the definition of macroscopic observables of the type (\ref{dens}).
As a rule, the substitution
\begin{equation}
\label{substitution}
   \int dp_{\parallel} \to eTE_0
\end{equation}
is introduced which results in the well known Schwinger formula \cite{Schwinger:1951nm} for the EPP production rate. 
The constant field model has been analyzed in detail in the recent work \cite{Tanji:2008ku}.

The strongly anisotropic nonequilibrium distribution (\ref{f_degen}) is defined by the symmetry 
of the external field and leads to strong longitudinal flow.
A detailed consideration of the nonequilibrium feature of this distribution can found in the work \cite{Spokoiny:1982pg}.

\begin{figure*}
\includegraphics[width=0.48\textwidth]{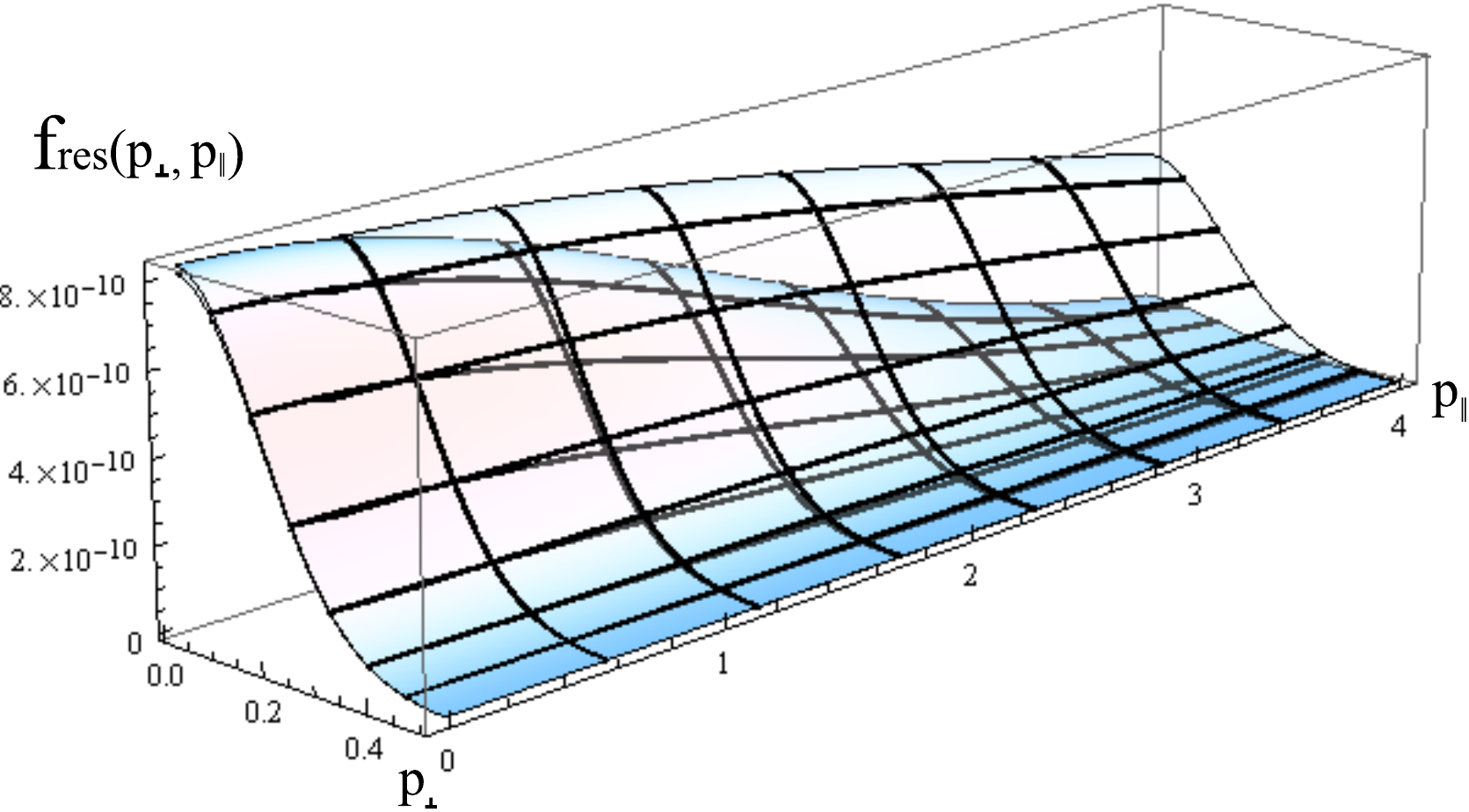} \hfill
\includegraphics[width=0.44\textwidth]{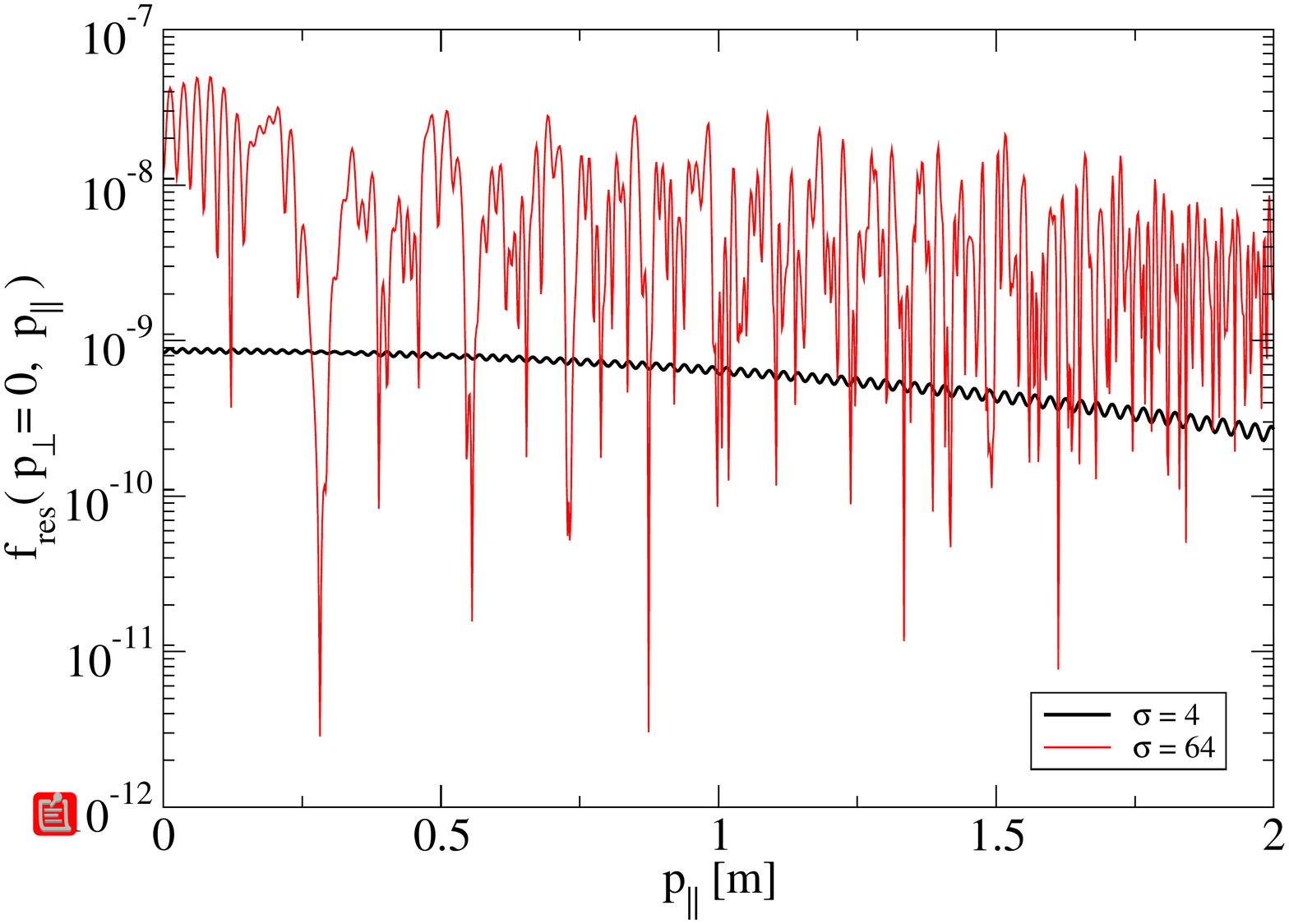}
\caption{
{\bf Left panel:} REPP distribution function $f(p_\bot,p_\parallel)$ for the Eckart-Sauter pulse (\ref{field2})  with $T = 82.4$ (bottom surface) and with $T = 164.8$ (top surface) at $E_0/E_c=0.15$ in momentum space for $0.0 \leq p_\bot \leq 0.5, 0.0 \leq   p_\parallel \leq 4.0$. {\bf Right panel:}  REPP distribution functions $f(p_\bot=0,p_\parallel)$ for a short pulse $\sigma = 4$ and a long pulse $\sigma = 64$ with the same amplitude of the field $E_0 = 0.15 E_c$. 
The cyclic frequency $\omega$ of the oscillating field satisfies the condition $1/\omega = 82.4$ that corresponds to the wavelength $0.1 nm$.\label{fig:6}}
\end{figure*}

The asymptotic distribution (\ref{f_degen}) in the constant field model is a smooth function of the transversal energy $\varepsilon_\bot(p_\bot)$.
In more realistic field models the structure of the distribution function becomes very complicated. 
As an example, see the right panel of Fig.~\ref{fig:6}.

\subsection{Non-monotonic entropy growth}

\begin{figure*}
\includegraphics[width=\textwidth]{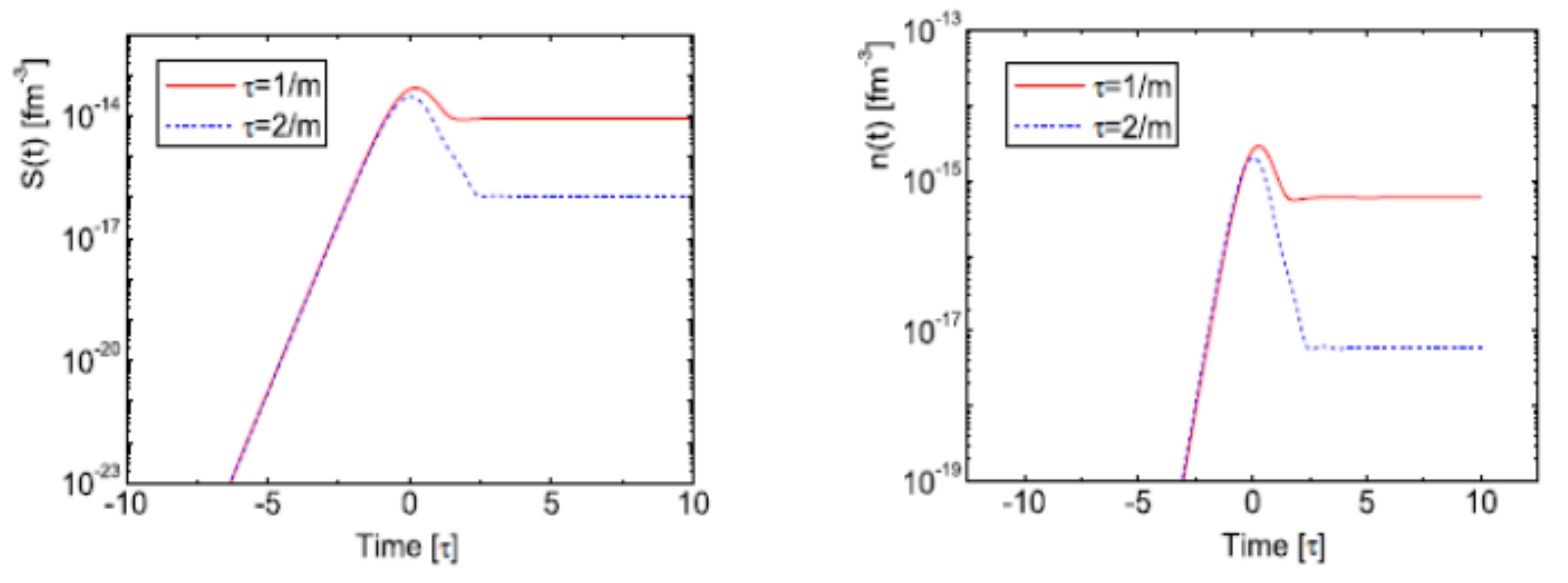}
\caption{Non-monotonous growth of the entropy (left panel) and of the density (right panel) for two values of the parameter $\tau$. 
\label{fig:8}}
\end{figure*}

The transition from the in-state to the out-state is accompanied by an non-monotonic entropy growth.
This phenomenon was marked and discussed long ego (e.g., \cite{Rau:1995ea,Cooper:1992hw, Habib:1995ee}).
For example, the function (\ref{f_degen}) leads to the following entropy production rate:
\begin{equation}\label{S_v}
\frac{S_{\rm out}}{T} = \frac{m^4}{8\pi^2} \frac{E_0}{E_c} \left( 1+ \frac{E_0}{\pi E_c }\right) 
\exp \left(-\pi \frac{E_c}{E_0} \right),
\end{equation}
where the pulse duration is defined by the relation (\ref{substitution}). 
In Eq.~(\ref{S_v}) the definition of the information entropy 
with the density 
\begin{equation}\label{S_d}
S(t) =  - \int \frac{d\mathbf{p}}{(2 \pi)^3} f(\mathbf{p} ,t) \ln{ f(\mathbf{p} ,t)}
\end{equation}
was used.
The most complete investigation was implemented  in the work \cite{Smolyansky_2012} on the basis of 
the KE (\ref{380}). Results from that work are shown in Fig.~\ref{fig:8}.

Let us notice that the KE  (\ref{380}) or the system of ODE (\ref{ode}) is not invariant with respect to time inversion.
Indeed, if $E(t)=E(-t)$ and $A(t)=-A(-t)$, one can see that replacement $f(\mathbf{p} ,t) \to f(-\mathbf{p} ,-t)$ is not change KE, but $f(\mathbf{p} ,t) \ne f(\mathbf{p} ,-t)$.
Thus, the entropy growth observed here is a result of  transforming of the primordial vacuum fluctuations under the action of a strong external field to the statistical ensemble of the EPP with well defined entropy.

\section{Schwinger process and thermalization}

As an application of the Schwinger process the particle production in heavy-ion collisions has been considered which may proceed by the decay of color electric flux tubes \cite{Casher:1978wy,Bialas:1984ye,Gatoff:1987uf}.
The flux tubes are characterized by a linear, stringlike potential between color charges, analogous to the case of a homogeneous electric field considered by Schwinger.
Using this analogy that $|eE|=\sigma$ with $\sigma \sim 0.19$ GeV$^2$ being the string tension, the 
transverse energy spectrum of produced particles according to the Schwinger mechanism would be 
(\ref{Schwinger})
\bea
\label{eq:Gaussian}
\frac{dN_{\rm Schwinger}}{d^2p_\perp} \sim \exp\left(- \frac{\pi \varepsilon_\perp^2}{\sigma} \right)~,
\eea
with $\varepsilon_\perp = \sqrt{m^2 + p^2_\bot}$ being the transverse energy (\ref{eq:energy_perp}),
often also denoted as "transverse mass" $m_\perp$.    
This spectrum of produced particles is nonthermal and thus would contradict the observation of thermal
particle spectra in heavy-ion collision experiments
\bea
\frac{dN_{\rm exp}}{d^2p_\perp} \sim \exp\left(- \frac{\varepsilon_\perp}{T_{\rm eff}} \right)~,
\eea
with an effective temperature $T_{\rm eff}\sim 180$ MeV (inverse slope parameter; see, e.g., \cite{Broniowski:2001we}). 
Thus the question for the thermalization arises. 
It has been suggested that it proceeds via collisions described by a kinetic equation 
\cite{Bialas:1984wv,Kajantie:1985jh}.
For a most recent discussion of the issue, see 
\cite{Ryblewski:2013eja,Gelis:2015kya,Gelis:2016rnt,Blaizot:2016iir}.
It has been questioned whether in high-energy nuclear collisions there is enough time for the thermal 
equilibration of the system by collisions, after the particle production in a Schwinger process.

As an alternative picture for the emergence of a thermal particle spectrum in ultrarelativistic particle collisions the analogue of the Hawking-Unruh radiation has been discussed 
\cite{Castorina:2007eb,Castorina:2014cia,Castorina:2014fna}.
This reasoning predicts thermal spectra of hadrons with the Hawking-Unruh temperature 
\bea
\label{eq:Hawking}
T_H = \sqrt{\frac{\sigma}{2\pi}}\sim 173~{\rm MeV}, 
\eea
where for the string tension $\sigma=0.19$ GeV$^2$ has been used.

In this context it is interesting to note a possible synthesis of both pictures as provided by the argument elucidated by Bialas \cite{Bialas:1999zg}.
If the string tension in the Schwinger process for flux tube decay would fluctuate and follow, e.g., a 
Poissonian distribution 
\bea
P(\sigma) = \exp(-\sigma/\sigma_0)/\sqrt{\pi\sigma\sigma_0}~,
\eea
which is normalized $\int d\sigma P(\sigma)=1$ and has a mean value 
$\langle \sigma\rangle=\int d\sigma \sigma P(\sigma)=\sigma_0/2$, 
then the initial Gaussian transverse energy spectrum (\ref{eq:Gaussian}) after averaging with the string tension fluctuations becomes exponential, i.e. thermal with the temperature parameter 
$T=\sqrt{\langle\sigma\rangle/(2\pi)}$,
\bea
\int d\sigma P(\sigma) \exp\left(- \frac{\pi \varepsilon_{\perp}^2}{\sigma} \right) 
= \exp\left(-\frac{\varepsilon_\perp}{T} \right)~.
\eea
Here the integral $\int_0^\infty dt \exp[-t - k^2/(4t)]/\sqrt{\pi t}=\exp(-k)$ has been used
\cite{Abramowicz}. 

This coincides with the Hawking-Unruh picture of thermal hadron production, where  in the case of fluctuating strings the string tension of Eq.~(\ref{eq:Hawking}) is now replaced by its mean value.   
We would like to note at this point that a largely thermal spectrum would arise also from the solution 
of a kinetic equation with the Schwinger source term, as has been demonstrated by Florkowski in 
Ref.~\cite{Florkowski:2003mm} for the case of parton creation (a more detailed calculation has recently
been done in  \cite{Ryblewski:2013eja}). This demonstrates the dynamical origin of thermal spectra. 

In order to draw the link to the observed hadron spectra in heavy-ion collision experiments, it remains to consider also the hadronization process when starting from the parton level of description.
For this purpose one could employ, e.g., kinetic theory approaches  built on the basis of the Nambu--Jona-Lasinio model Lagrangian, see  \cite{Rehberg:1995kh,Rehberg:1998me,Friesen:2013bta,Marty:2015iwa}.
In this context the dynamical chiral symmetry breaking in the quark sector plays an essential role as it triggers the binding of quarks into hadrons (inverse Mott effect). The increase in the sigma meson mass that accompanies the dynamical chiral symmetry breaking gives rise to additional sigma meson production by the inertial mechanism  (see \cite{Juchnowski:2015uqj} and references therein). 
By the dominant decay $\sigma\to\pi\pi$ this leads to an additional population of low-momentum pion states and can contribute to the observed effect that s also discussed as a precursor of pion Bose condensation and may simultaneously resolve the LHC proton puzzle \cite{Begun:2013nga} within a non-equilibrium model.

\section{Summary \label{sect:3}}

In these lectures, we have given the derivation of the kinetic equation for pair production in a homogeneous, time-dependent external field (dynamical Schwinger mechanism) and discussed the  Markovian as well as the low-density limiting cases.
Further, we have investigated numerical solutions of the full kinetic equation for one-sheeted and multi-sheeted external field models. 
We have demonstrated that the dynamical Schwinger mechanism can be considered as a field induced phase transition from the primordial in-state with latent vacuum fluctuations to the final massive quantum field system of particle-antiparticle pairs under the action of a strong external field. 
Thus, in the simplest situation, this process starts from the vacuum in-state (in the considered statement of the problem the electrons and positrons are absent in the initial state) and possesses the following characteristic features:
\begin{enumerate}
\item[(i)] presence of three different stages in the time evolution: smooth quasiparticle, fastly oscillating transient and asymptotic final stage;
\item[(ii)] strong nonequilibrium character of FIPT, including the out-state;
\item[(iii)] non-monotonic entropy growth.
\end{enumerate}

The nonperturbative kinetic description is a reliable basis for the investigation of secondary physical phenomena such as, for example, the electromagnetic radiation from the region of the action of an external field \cite{Blaschke:2013ip}.
Here one can expect that the considered stages of the FIPT will be reflected in the time-based sweeps of these phenomena.

On the other hand, these features are rather universal and are characteristic on the qualitative level for physical systems of different nature.

On the formal level this universality appears because the corresponding KE's belong to the united class of integro-differential equations of non-Markovian type with fastly oscillating kernel.
Examples of this kind are, e.g., KE's for description of the vacuum creation of scalar bosons and of fermions in the FRW space-time \cite{Grib:1994}, the noncontradictory KE for massive vector bosons in the same metric space \cite{Grib:1994} and the nonperturbative KE for description of the carrier excitations in graphene \cite{Panferov:2017}.

In these lectures we have restricted ourselves to the consideration of the domain of the tunneling mechanism of particle creation, $\gamma \ll 1$. 
The first step in investigation of the few-photon domain of EPP creation ($\gamma \gg 1$) 
has been made in the work \cite{Panferov_2017_2}.

We have discussed the application of the Schwinger process in the description of particle production in heavy-ion collisions, mentioning the initial ideas as well as recent developments devoted to the understanding of the apparent thermal spectra of produced hadrons as well as the recent pion and proton puzzles observed at CERN in LHC experiments. 
The kinetic approach to particle production in strong, time-dependent fields is indispensable for a microphysical elucidation of the mechanisms at work under these extreme conditions. 

\subsection*{Acknowledgements}
We thank A. Fedotov, T. Fischer, W. Florkowski, D.M. Gitman, B. K\"ampfer, A. Otto, R. Ryblewski, D. Seipt and A. Titov for discussions and collaboration. 
We acknowledge support by the HISS Dubna program for our participation at this Summer School, 
by NCN grant No. UMO-2014/15/B/ST2/03752 (D.B., J.L., A.S.), by COST Action CA15213 "THOR" (D.B., L.J.) and by IFT internal grant No. 1441/M/IFT/15 (L.J.).    


\begin{footnotesize}



%

\end{footnotesize}


\end{document}